\documentclass[showpacs,pra,superscriptaddress,twocolumn]{revtex4-1}
\usepackage{pgf}
\usepackage[utf8]{inputenc}
\usepackage{amsmath}
\usepackage{braket}
\usepackage{amssymb}
\usepackage{amsmath}
\usepackage{ulem}
\usepackage{stackrel}
\usepackage{enumitem} 
\usepackage{appendix}
\usepackage{graphicx}
\usepackage{overpic}
\usepackage{epstopdf}
\usepackage[pdftex,bookmarks,colorlinks]{hyperref}
\usepackage{hyperref}
\usepackage{amsmath}
\usepackage{amssymb}
\usepackage{braket}
\usepackage{ulem}

\newcommand{\Tr}[1]{\operatorname{Tr} #1}
\newcommand{\mean}[1]{\left\langle #1\right\rangle}
\newcommand{\nep}{\textrm{e}}

\definecolor{caribbeangreen}{rgb}{0.0, 0.8, 0.6}

\begin{document}
\title{Quantum chaos and ensemble inequivalence of quantum long-range Ising chains}
%
\author{Angelo Russomanno}
\affiliation{Max-Planck-Institut f\"ur Physik Komplexer Systeme, N\"othnitzer Stra{\ss}e 38, D-01187, Dresden, Germany}
\author{Michele Fava}
\affiliation{Rudolf Peierls Centre for Theoretical Physics, Clarendon Laboratory, University of Oxford, Oxford OX1 3PU, United Kingdom}
\author{Markus Heyl}
\affiliation{Max-Planck-Institut f\"ur Physik Komplexer Systeme, N\"othnitzer Stra{\ss}e 38, D-01187, Dresden, Germany}
\normalem
\begin{abstract}
We use large-scale exact diagonalization to study the quantum Ising chain in a transverse field with long-range power-law interactions decaying with exponent $\alpha$. We numerically study various probes for quantum chaos and eigenstate thermalization {on} the level of eigenvalues and eigenstates. The level-spacing statistics yields a clear sign towards a Wigner-Dyson distribution and therefore towards quantum chaos across all values of $\alpha>0$. Yet, for $\alpha<1$ we find that the microcanonical entropy is nonconvex. This is due to the fact that the spectrum is organized in energetically separated multiplets for $\alpha<1$. {While quantum chaotic behaviour develops within the individual multiplets, many multiplets don't overlap and don't mix with each other}, as we analytically and numerically argue. {Our findings suggest that a small fraction of the multiplets could persist at low energies for $\alpha\ll 1$ even for large $N$, giving rise to ensemble inequivalence.}

\end{abstract}
\maketitle
\section{Introduction}

Thermalization in classical Hamiltonian systems is well understood in terms of chaotic dynamics and the related essentially ergodic exploration of the phase space~\cite{lichtenberg1983regular,Vulpiani,Berry_regirr78:proceeding}. From the quantum point of view the
physical mechanism is quite different, {with} the eigenstates of the Hamiltonian {behaving similar to the eigenstates of a random matrix with the additional property that }
they appear
thermal from the point of view of local measurements. This is the paradigm of eigenstate thermalization (ETH) introduced in Refs.~\cite{Deutsch_PRA91,Sred_PRE94,Rigol_Nat,Prosen_PRE99}. In general there is correspondence between classical and quantum thermalization~\cite{Berry1_1977,PhysRevLett.51.943,Sred_PRE94,PhysRevA.34.591,Prosen_AJ,PhysRevLett.52.1,PhysRevLett.75.2300}, but due to the different physical mechanism there can be cases where quantization breaks ergodicity, as {for} many-body localization (see~\cite{Bloch_2019} for a review) and many-body dynamical localization~\cite{QRKR,kicked-rotors,rylands,Michele_arxiv,Prosen_PRE99,Prosen_PRL98}.

In quantum short-range thermalizing systems there are three strictly related properties. First of all eigenstate thermalization, that's to say that almost all the excited eigenstate{s} locally {behave equal to the} microcanonical {or thermal} density matrix~\cite{Kafri}. So, expectation {value}s of local observables equal the corresponding microcanonical ones, up to fluctuations vanishing in the thermodynamic limit. This property is strictly related to a second one{:} quantum chaos~\cite{Kafri}. Quantum chaos means that the {spectrum of the Hamiltonian} behaves essentially as {the one of} a random matrix~\cite{Haake} and this occurs typically for many-body nonintegrable models~\cite{Poil} and for Hamiltonians obtained quantizing classical chaotic systems~\cite{PhysRevLett.52.1}. 
Hamiltonians {show in general eigenstate thermalization together with quantum chaos} and behave as random matrices~\cite{Kafri} (with some caveats~\cite{Ivan2020}).  This fact gives rise to random eigenstates which look locally thermal as appropriate for ETH. A third property relevant in thermalized short-range interacting systems is additivity and ensemble equivalence which are strictly related to a convex microcanonical entropy~\cite{Mukamel2}.


An interesting question is if {what is the relation} between quantum chaos, ETH and ensemble equivalence in quantum systems with long-range interactions. In the classical case, for instance, the thermalization behavior is very different in the case of short- and long-range interactions. For classical systems with short-range interactions, any nonlinear Hamiltonian with more than two degrees of freedom and no conservation law beyond energy gives rise to chaos, {essentially ergodic dynamics}~\cite{lichtenberg1983regular} and ensemble equivalence~\cite{Vulpiani}. In the long-range case the situation is very different. {A central aspect of long-range classical systems is the inequivalence of canonical and microcanonical ensemble due to the lack of additivity of the Hamiltonian~\cite{Antoni,Mukamel1,Mukamel2}. This implies that the dynamics does not lead to a simple thermalization behavior, even in presence of chaos. One can see an effectively regular} behavior dominated by one or few degrees of freedom~\cite{Antoni,Firpo,latora,anteneodo,rocha,long_range_book} which has been exploited to obtain a classical Hamiltonian time crystal~\cite{PhysRevLett.123.184301}. 

{Although ensemble inequivalence for the exactly-solvable infinite-range anisotropic quantum Heisenberg
model has been studied in~\cite{Kast1,Kast2,Kast3},} the relation between quantum chaos and ensemble equivalence in generic interacting quantum long-range systems has not yet been explored. 
We fill here this gap focusing on a long-range ferromagnetic Ising spin-$1/2$ chain model. Similar models have been already studied. One very well studied case is the Ising model with infinite-range interactions (the so called Lipkin-Meshkov-Glick model) which is known to be integrable~\cite{Lipkin_NucPhys65,PhysRevB.86.184303,Sciolla_2}. It is also known that the isotropic Heisenberg chain with power-law interactions with exponent $\alpha=2$ is integrable~\cite{haldane_3,10.1007/978-3-642-85129-2_1} as well as some anisotropic spin-chain models with $\alpha=2$~\cite{uglov1995trigonometric,SECHIN20181,PhysRevB.97.214416}. Spin chains with disorder and power-law interactions are known to undergo a transition between a many-body-localized-like and an ergodic phase~\cite{Hauke2015,Smith_2016,Burin_2017,Roy_2019,PhysRevLett.113.243002,Tikhonov,PhysRevB.91.094202}. 

Comparatively less attention has been devoted to {homogeneous} long-range interacting spin models. Although these models have been extensively studied in the context of quantum quenches~\cite{Piccitto_2019,Silva,PhysRevB.99.045128,PhysRevB.98.134303,2017Halimeh,
PhysRevLett.122.150601,guo2019signaling,PhysRevB.99.121112,verdel2019realtime,Silva,reyhaneh,PhysRevA.99.010105,Hauke_2013,Colmenarez_2020} and quantum spin liquids~\cite{chiocchetta2020cavityinduced}, {and their dynamics has attracted a lot of experimental interest~\cite{neyenhuis2016observation,tan2019observation,monroe2020programmable,Zhang_2017,Jurcevic_2017,Brydges_2019,Smith_2016,Lanyon_2017},} an analysis of the thermalization properties of the eigenstates is generally lacking. {A significant exception is~\cite{fratus2017eigenstate} which showed {quantum chaos} at low energies for $\alpha=1.5$ in the clean ferromagnetic spin-$1/2$ Ising model with long-range power-law interactions.} The dynamics of this model has been intensively studied, mostly in connection with the persistence of long-range order in the asymptotic state of the dynamics~\cite{Piccitto_2019,Sciolla_2,Silva,PhysRevB.99.045128,PhysRevB.98.134303,
PhysRevLett.122.150601,guo2019signaling,PhysRevB.99.121112,verdel2019realtime,Silva,reyhaneh}, for different values of $\alpha$ and small transverse field, but it is not known if this asymptotic state is thermal.

In our work we {focus on this same model and widely extend the ETH and quantum chaos analysis by using exact diagonalization and exploring a wide range of $\alpha$ and energies.}
The main question {is} {the relation between} eigenstate thermalization, quantum chaos and convex microcanonical entropy. 
For $\alpha<1$ we find  a very interesting behavior. 

On the one hand the level spacing statistics gives a clear answer pointing towards a random-matrix Wigner-Dyson form. {This is valid} for any value of $0<\alpha<\infty$, but for the region around $\alpha \approx 2$ for weak transverse fields around,  hinting to possible the vicinity of some integrable point. 

On the other hand ETH indicators (eigenstate expectations and eigenstate half-system entanglement entropies) yield a much less clear perspective for finite system sizes, in particular, for $\alpha<1$.
We find that the permutation symmetry, which is only exact at $\alpha=0$, leaves behind a strong fingerprint in many ETH indicators: The  $\alpha=0$ symmetry-protected multiplets in the energy spectrum represent a relatively rigid structure for $0<\alpha<1$. They affect the eigenstate quantities and forbid them a smooth ETH dependence on energy, in contrast with short-range interacting quantum-chaotic systems~\cite{Kafri}. 

These multiplets have another important consequence: The microcanonical entropy {becomes} a nonconvex function of energy, which in the thermodynamic limit excludes ensemble equivalence in a thermodynamic sense. {We provide an analytical argument for the rigidity of the multiplets for large but finite $N$} when $\alpha<1$. {For $\alpha\ll 1$ we observe that {some of} the multiplets { at low energies persist also for large $N$. }}{As a consequence, we argue that the system doesn't obey} ensemble equivalence. 

These observations {on the multiplet structure} seem to contradict our findings for the level spacing statistics. {These results are reconciled} by {what we call a} partial spectral quantum chaos.
The states in individual multiplets, which are separated in energy with respect to each other, mix in a quantum chaotic fashion, whereas the multiplets don't yet mix among each other for the accessible system sizes. Each multiplet in the bulk of the spectrum behaves as a separate random matrix leading to a overall Wigner-Dyson level statistics. {This is a significant result: each multiplet behaves a random matrix from a spectral point of view, so its spectrum {tends to a smooth continuum for $N\to\infty$}. This is in contrast {to} integrable long-range system whose spectrum has been claimed to be pure point also in the thermodynamic limit~\cite{Fenu}.} 

We emphasize again that we expect the multiplet structure to {be most rigid} at low energy densities, which might have important consequences for the absence of thermalization observed in low-energy quenches~\cite{Silva,reyhaneh}.

The paper is organized as follows. 
In Sec.~\ref{model:sec} we define the model Hamiltonian. 
In Sec.~\ref{spectrum:sec} we study the quantum chaos properties at the level of the spectrum. We show a generalized tendency towards a Wigner-Dyson level-spacing statistics for increasing system size. 
In Sec.~\ref{randma:sec} we discuss an analytical argument based on the random-matrix behavior of each multiplet. We show that the spectral multiplet width increases linearly in $\alpha$, in agreement with numerics, and that part of the multiplets {persist} in the large-$N$ limit, {for low energies and} $\alpha\ll 1$.
In Sec.~\ref{multiplets:sec} we better discuss the multiplet spectral structure for small $\alpha$ and finite $N$ and study the corresponding nonconvex behavior of the microcanonical entropy related to ensemble inequivalence. {In Sec.~\ref{pairing:sec} we study the broken symmetry edge (the energy density below which there is $\mathbb{Z}_2$ symmetry breaking) and find a different behavior in the canonical and microcanonical ensemble, although there are too strong finite-size effects to allow to make statements on ensemble inequivalence. }
We study also the eigenstate properties by considering the eigenstate expectation {values} of a local operator, the longitudinal nearest-neighbour correlation (Sec.~\ref{nearco:sec}), and of the half-system entanglement entropies of the eigenstates (Appendix~\ref{entropies:sec}). 

 {In Appendix~\ref{hilbers:sec} we discuss the Hilbert-Schmidt distance of the $\alpha>0$ Hamiltonian from the $\alpha=0$ Hamiltonian, showing its linearity in the limit $\alpha\to 0$. This fact, together with the random-matrix assumption, allows us to explain the linearity in $\alpha$ of the multiplet spectral width in Sec.~\ref{randma:sec}.
%

\section{Model Hamiltonian} \label{model:sec}
In this work we study the ferromagnetic long-range interacting quantum Ising chain in a transverse field:
\begin{equation} \label{model:eqn}
  \hat{H}^{(\alpha)}=-\frac{J}{N(\alpha)}\sum_{i,j,\,i\neq j}^N\frac{\hat{\sigma}_i^z\hat{\sigma}_j^z}{D_{i,\,j}^\alpha}+h\sum_{i=1}^N \hat{\sigma}_i^x\,.
\end{equation}
Here, $\sigma_i^{\alpha}$ with $\alpha=x,y,z$ denotes the Pauli matrices at lattice site $i=1,\dots,N$ with $N$ the system size.
We use periodic boundary conditions implemented through the definition~\cite{PhysRevLett.122.150601} $D_{i,\,j}\equiv\min[|i-j|,N-|i-j|]$; we define the Kac factor~\cite{kac} $N(\alpha)\equiv\frac{1}{N-1}\sum_{i,j,\,i\neq j}\frac{1}{D_{i,\,j}^\alpha}$ in order to preserve extensivity of the Hamiltonian. 

We use exact diagonalization. We largely exploit the translation, inversion and $\mathbb{Z}_2$ ($\hat{\sigma}_i^z\to-\hat{\sigma}_i^z$) symmetries of the model in order to restrict to an invariant subspace of the Hamiltonian. In most of the text we restrict to the subspace fully symmetric under all the symmetries of the Hamiltonian. We call this Hamiltonian eigenspace $\mathcal{H}_S$ and 
we define it as the zero-momentum sector subspace even with respect to inversion and $\mathbb{Z}_2$ symmetry. For future convenience we define $\mathcal N_S\equiv \dim \mathcal{H}_S$. In Sec.~\ref{pairing:sec} we are interested in the spectral pairing properties of the model{, which requires to consider both $\mathbb{Z}_2$ symmetry sectors}: {We consider here} the zero-momentum sector subspace even only with respect to inversion. 
We {denote} the eigenstates of the Hamiltonian $\ket{\varphi_\mu}$ and the corresponding eigenenergies $E_\mu$ (taken in increasing order), {while always} specifying which subspace we are considering.

In the limit $\alpha\to\infty$ the model in Eq.~\eqref{model:eqn} reduces to the nearest-neighbour quantum Ising chain. This model is integrable and undergoes a quantum phase transition: Its ground state breaks the $\mathbb{Z}_2$ symmetry for $h<1$~\cite{Sachdev,mbeng2020quantum}. For any finite system size, the ground state is {doubly degenerate made up by the two states symmetric and antisymmetric under the global $\mathbb{Z}_2$ symmetry}, with a splitting exponentially small in the system size. The states in the doublet show long-range order and  the doublet becomes degenerate in the thermodynamic limit, giving rise to symmetry breaking. 

In the limit $\alpha= 0$, on the opposite, Eq.~\eqref{model:eqn} reduces to the Lipkin-Meshkov-Glick model. This model is also integrable, thanks to the full permutation symmetry, and it shows a symmetry-broken phase for $h<1$. In contrast {to} the $\alpha\to\infty$ case, all the spectrum up to an extensive energy $Ne^*$ is organized in doublets with exponentially small splitting and the corresponding eigenstates have long range order~\cite{Lipkin_NucPhys65,PhysRevB.86.184303,Sciolla_2}. {Due to the full permutation symmetry, the Hilbert space is factorized in a number of invariant subspaces, differently transforming under the permutation symmetries~\cite{PhysRevB.86.184303}. The number of these subspaces is exponential in $N$, and many of them have the same level structure. This gives rise to massively degenerate multiplets, whose levels belong to different symmetry sectors, a property which will be quite relevant in the following.}

{For $\alpha=0$,} {the number of distinct multiplets is set by} the possible {distinct} simultaneous eigenstates of the square total spin $\hat{S}^2=\frac{1}{4}(\sum_j\hat{\boldsymbol{\sigma}})^2$ and the total spin $z$ component $\hat{S}^z=\frac{1}{2}\sum_j\sigma_j^z$. This is a consequence of the total-spin conservation and the
{ {permutation} symmetry of the Hamiltonian}~\cite{Lipkin_NucPhys65}. The square total spin has eigenvalues $S(S+1)$ with $S$ going from $S=0$ to $S=N/2$ and for each value of $S$ the total $z$ component can acquire $2S+1$ values. Assuming $N$ from now on even -- so that $S$ assumes only integer values -- the number of multiplets is $\mathcal{Q}=\sum_{S=0}^{N/2}(2S+1)=(N/2+1)^2$. For $\alpha=0$ each multiplet is degenerate with degeneracy $g(S)$ given only by $S$ and $N$ through the formula~\cite{PhysRevB.86.184303}
\begin{equation}\label{gigi:eqn}
  g(S) = \binom{N}{\frac{N}{2}+S} - \binom{N}{\frac{N}{2}+S+1}
\end{equation}

In the remainder of the paper we consider the case of intermediate $\alpha$. 

%
%
%

\section{Quantum chaos and level spacing statistics} \label{spectrum:sec}
First, we study the quantum chaos properties focusing on the level spacing statistics. 
The model in Eq.~\eqref{model:eqn} is integrable for the limits $\alpha=0$ (infinite-range case) and $\alpha\to\infty$ (nearest-neighbour case). We {now} aim at exploring the behavior at intermediate $\alpha$. 

{For concreteness, we don't scan extensively across the transverse fields, but rather focus on two representative values} $h=0.1$ and $h=0.5$. {In Fig.~\ref{correl1:fig} we investigate the spectral properties of the model as a function of $\alpha$ upon varying the system size $N$.} {Specifically, we plot the} average level spacing ratio, $r$ (introduced in~\cite{Pal_PhysRevB10}), {which is a central probe for quantum chaos}  and is defined as
\begin{equation} \label{rorro:eqn}
  r=\frac{1}{\mathcal{N}_S-2}\sum_{\mu=1}^{\mathcal{N}_S-2}\frac{\min(E_{\mu+2}-E_{\mu+1},E_{\mu+1}-E_\mu)}{\max(E_{\mu+2}-E_{\mu+1},E_{\mu+1}-E_\mu)}\,.
\end{equation}
%
With the time-reversal symmetry properties of our Hamiltonian, a value $r=r_{\rm WD}\simeq 0.5295$ would be associated {with} a fully quantum-chaotic random-matrix-like behavior {given by the Gaussian Orthogonal Ensemble (GOE)} and a Wigner-Dyson distribution for the level spacings~\cite{Haake}. On the opposite, a value $r=r_P\simeq 0.386$ is known to be related to a Poisson distribution of the level spacings, which implies integrable behavior~\cite{Berry_PRS77}.

Before considering the behavior for large $\alpha$ (Sec.~\ref{alpha_grande:sec}) and $\alpha\ll 1$ (Sec.~\ref{alpha_piccolo:sec}), and the associated tendency towards quantum chaos for increasing $N$, let us say something about the 
strong minimum at $\alpha=2$ appearing in Fig.~\ref{correl1:fig}(a). It suggests a behavior {closer} to integrability (and the corresponding Poisson value) which persists at least up to $N=22$. It is important to remind that there are spin models with power-law interactions decaying with $\alpha=2$ that are integrable, such as the long-range isotropic Heisenberg chain~\cite{haldane_3} or other anisotropic long-range models~\cite{uglov1995trigonometric,SECHIN20181,PhysRevB.97.214416}. 
{It could be an interesting question for future research} to {investigate} {whether} this phenomenon is related to the proximity to an integrable point. 

%
\begin{figure}
  \begin{center}
   \begin{tabular}{c}
%
      \begin{overpic}[width=80mm]{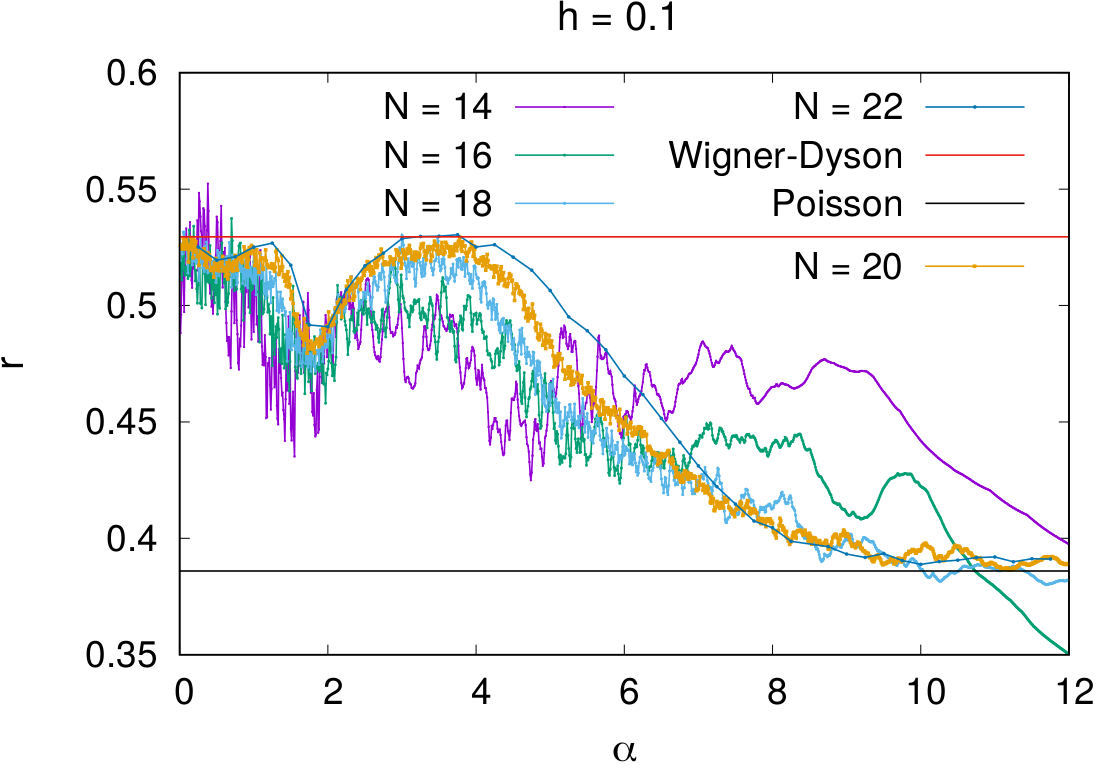}\put(90,36){(a)}\end{overpic}\\
      \vspace{0.1cm}\\
      \begin{overpic}[width=80mm]{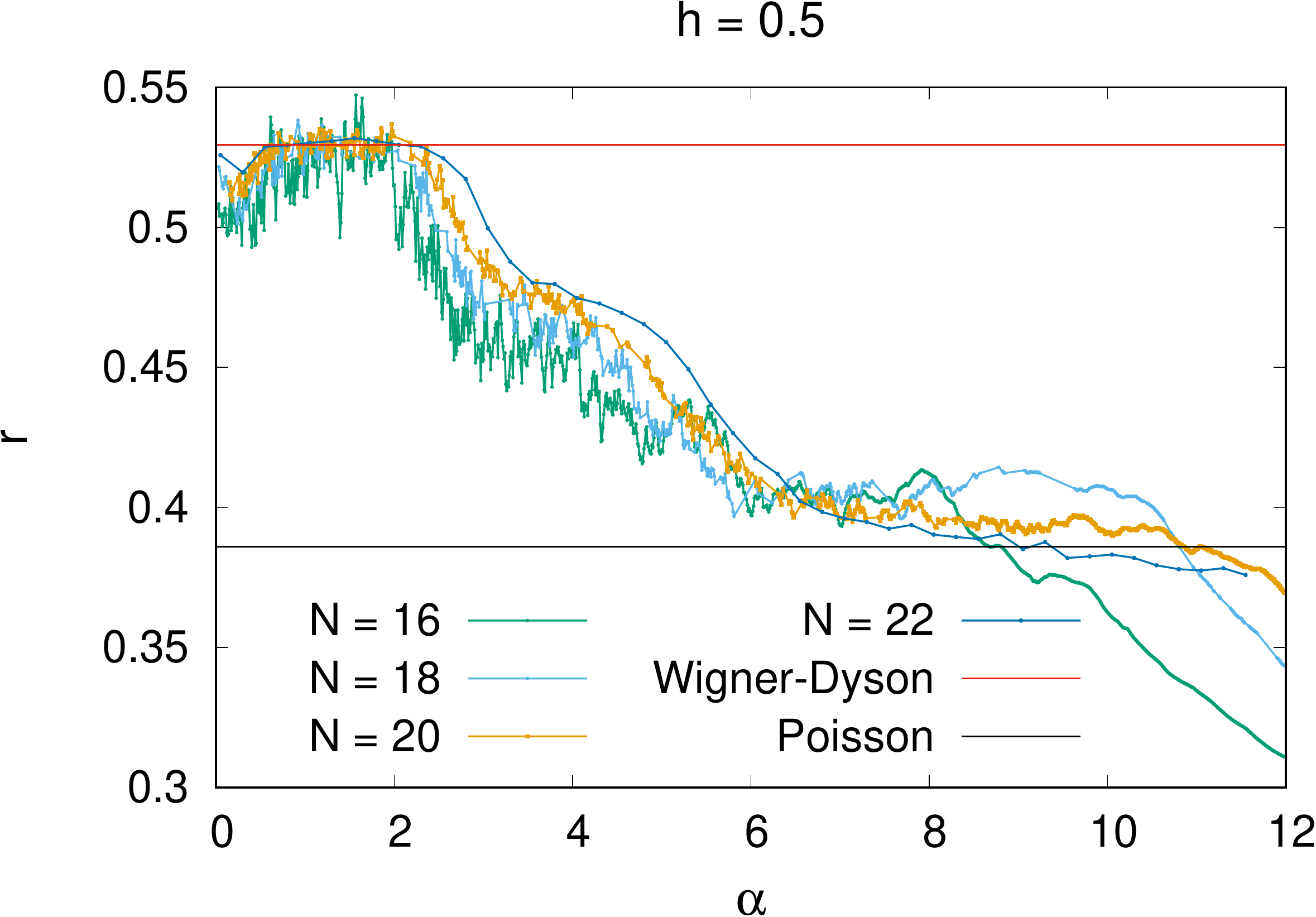}\put(90,41){(b)}\end{overpic}
   \end{tabular}
  \end{center}
 \caption{Average level spacing ratio versus $\alpha$. We consider $h=0.1$ [panel~(a)] and $h=0.5$ [panel~(b)].}
    \label{correl1:fig}
\end{figure}

\subsection{Large $\alpha$} \label{alpha_grande:sec}
For large $\alpha$ we see {in Fig.~\ref{correl1:fig} that {there is a crossover} towards} the Poisson value $r_P$. At some {larger} value of $\alpha$ there is {another} {crossover towards a value even smaller than Poisson}. This behavior of $r$ is a {finite-size effect due to the proximity }of the integrable $\alpha\to\infty$ point. The spectrum becomes quantum chaotic in the thermodynamic limit: {As we are going to show}, the crossover towards Poisson shifts to large $\alpha$ for increasing $N$.

We {can argue this shift towards integrability as follows.}
{In a free-fermion model (corresponding to our $\alpha \to \infty$ case), any arbitrarily small integrability-breaking next-nearest-neighbour interaction restores thermalization in the thermodynamic limit~\cite{PhysRevLett.115.180601, PhysRevB.94.245117}. Similarly, in our case, for $\alpha\gg 1$, the next-nearest neighbour terms are the stronger ones breaking the integrability of the nearest-neighbour $\alpha\to\infty$ model. {For increasing $N$, the {next-}nearest-neighbour terms become at some point large enough compared to the level spacings, and the model becomes {quantum chaotic}}}

{{Let us now} roughly estimate {the crossover scale} at which the system becomes quantum chaotic for $\alpha\gg 1$, {by comparing} the next-nearest neighbour interaction term with the relevant gap $\Delta$ of the integrable nearest-neighbour model.} The next-nearest neighbour term is of order $V\sim J / (N(\alpha) 2^\alpha)$. We can understand the relevant gap of the nearest-neighbour model, moving to its fermionic representation via the Jordan-Wigner transformation~\cite{LIEB1961407}. In this representation, the nearest-neighbour model is integrable and its excitations are fermionic quasiparticles~\cite{Pfeuty,mbeng2020quantum} with energy $\epsilon_k={2}\sqrt{\left(\frac{J}{{N(\alpha)}}\right)^2+h^2-2\frac{J}{{N(\alpha)}}h\cos k}$. We have $k\in[0,\pi]$ and, for finite system size $N$, $k$ can take only $N$ discrete equally spaced values. In the fermionic representation the next-nearest-neighbour term becomes a four-fermion term which induces inelastic scattering between the fermionic quasiparticles. If momenta $k_1$ and $k_2$ go into momenta $k_3$, $k_1+k_2-k_3$, the relevant gap is $\Delta = \epsilon_{k_3}+\epsilon_{k_1+k_2-k_3}-\epsilon_{k_1}-\epsilon_{k_2}$. We can roughly estimate $\Delta$ by taking twice the bandwidth of $\epsilon_k$ and dividing it by $N$, the number of allowed equally-spaced $k$ values. We find
\begin{equation}
 \Delta\sim \frac{4}{N}\left[\left|\frac{J}{N(\alpha)}+h\right|-\left|\frac{J}{N(\alpha)}-h\right|\right]\,.
\end{equation}
Imposing that $V\gtrsim \Delta$, one finds that quantum chaotic behavior is obeyed for $\alpha\lesssim \alpha^*$. 
%
%
We evaluate $\alpha^*$ numerically, and find that $\alpha^*$ asymptotically increases as $\log_2N$ (see Fig.~\ref{crossing:fig}). So, for $N\to\infty$ there is quantum chaos for all values of $h$. 
\begin{figure}
  \begin{center}
   \begin{tabular}{c}
%
      \begin{overpic}[width=80mm]{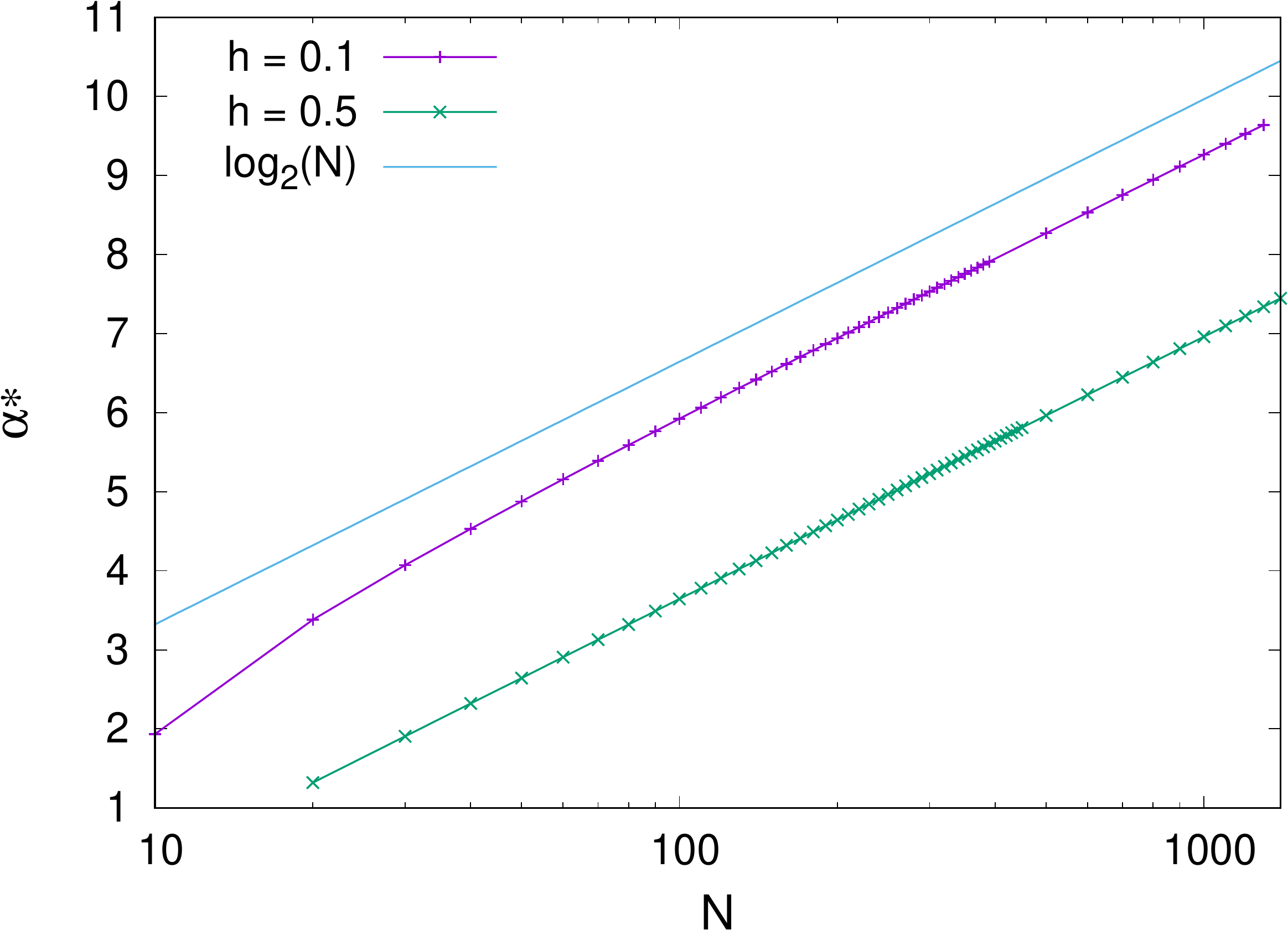}\put(90,36){}\end{overpic}\\
   \end{tabular}
  \end{center}
 \caption{$\alpha^*$ versus $N$ (single-logarithmic plot). Notice the increase as $\log_2N$.}
    \label{crossing:fig}
\end{figure}
%
\subsection{The role of multiplets for $\alpha\ll 1$} \label{alpha_piccolo:sec}
For $\alpha\ll 1$ $r$ is {close} to the Wigner-Dyson value (Fig.~\ref{correl1:fig}).   {Therefore, {our numerics suggests that} the integrable behavior at $\alpha=0$~\cite{Lipkin_NucPhys65} is unstable to a small perturbation in $\alpha$ which breaks the full permutation symmetry at $\alpha=0$.} 

{As we have already discussed in Sec.~\ref{model:sec}, the multiplets at $\alpha=0$} do not correspond to a given permutation symmetry class, but contain states belonging to different invariant subspaces, differently transforming under permutation. There are many subspaces with the same energy levels inside~\cite{PhysRevB.86.184303}. When perturbation symmetry is broken by $\alpha\ll 1$, the degenerate states inside each multiplet can mix and so all the subspaces are mixed by the Hamiltonian. This leads to quantum chaos, {as we are going to argue}. 

Since there is no gap to protect the {subspaces} from mixing, this change happens abruptly as soon as $\alpha>0$ and the multiplet degeneracy is lifted. We can see an example of that in Fig.~\ref{plot_multi:fig}. We plot $E_\mu$ versus $\mu/\mathcal{N}_S$ for $h=0.1$ and two values of $\alpha$, $\alpha=0$ and $\alpha=0.15$. For $\alpha=0$ there are many degenerate multiplets at all energies, as we can see in the magnifying insets. For $\alpha=0.15$ the multiplets merge into a smooth continuum at large energy (right inset) but {can be} still {well identified} at low energy (left inset). The organization of the spectrum in multiplets for small $\alpha$ is also evident in the eigenstate expectation of local observables (Sec.~\ref{nearco:sec}) and the half-system entanglement entropy of these eigenstates (Appendix~\ref{entropies:sec}).

{This multiplet structure is apparently in contrast with the average level spacing ratio being close to the Wigner-Dyson value. In order to explain this apparent contradiction, we notice that the number of  gaps among multiplets} is much {smaller} than the {total} number of states. The number of discontinuity points scales as the number of distinct multiplets at $\alpha=0$, which scales as $N(N+1)/2$ (see Sec.~\ref{multiplets:sec}), while the number of states equals $\mathcal{N}_S$ which is exponential in $N$. So, if each of the multiplets behaves separately as a random matrix, the overall average level spacing ratio is Wigner Dyson in the large $N$ limit. {This is exactly what happens, as we show in detail in the next section.}
\begin{figure}
  \begin{center}
   \begin{tabular}{c}
     \includegraphics[width=8cm]{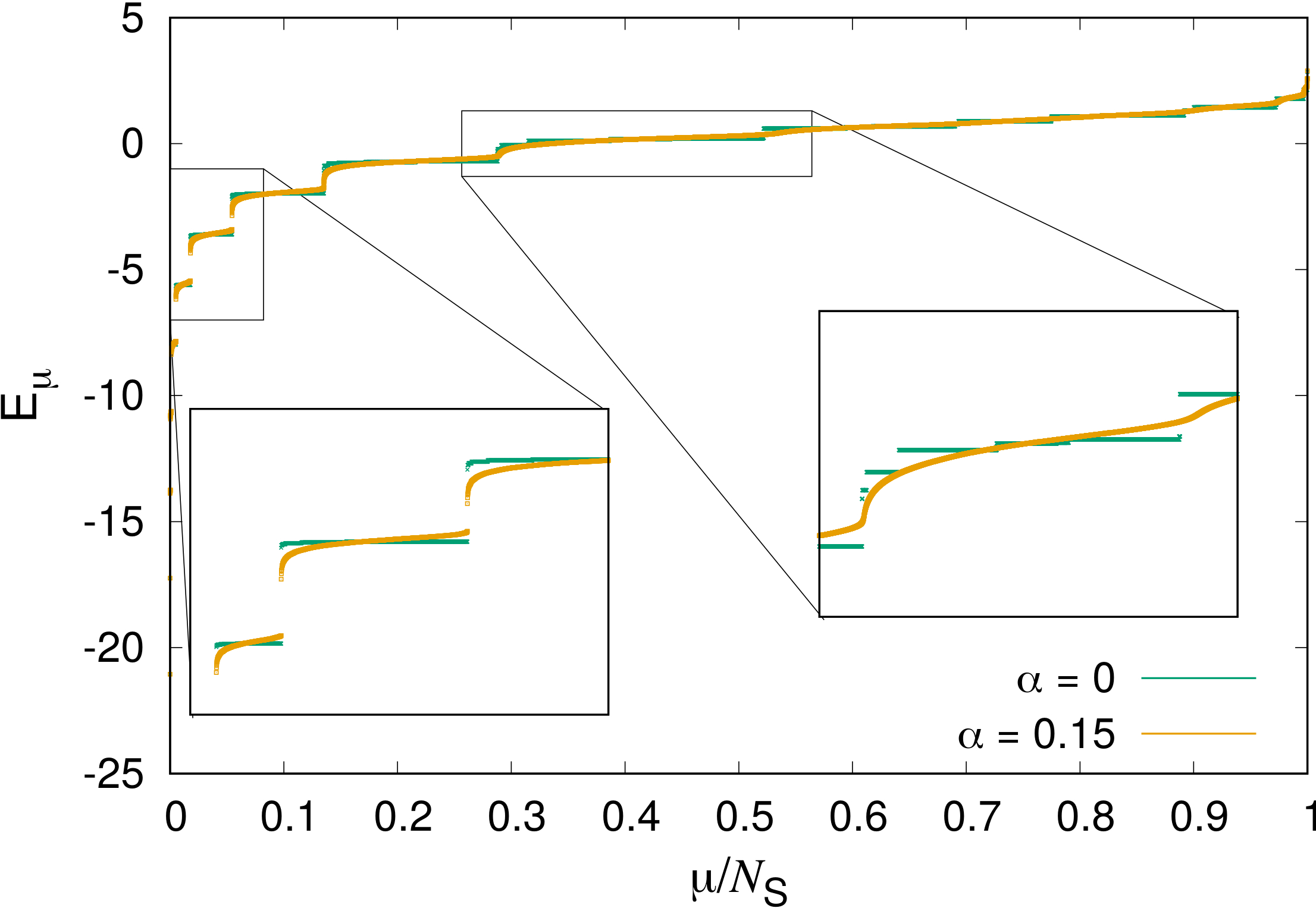}\\
   \end{tabular}
  \end{center}
 \caption{Plot of $E_\mu$ versus $\mu/\mathcal{N}_S$ for $h=0.1$, $N=22$ and two different values of $\alpha$.}
    \label{plot_multi:fig}
\end{figure}


%


\section{Random-matrix behavior and  multiplet spectral width for $\alpha<1$} \label{randma:sec}
{Goal of this section is to argue that} each multiplet broadens by an amount proportional to $\alpha$. This numerically verified statement relies on the Hamiltonian projected to a multiplet subspace behaving like a GOE random matrix, as we argue in Sec.~\ref{wsm:sec}. {The main implication is that the total multiplet width is linear in $N$ and much smaller than the total spectral width for $\alpha\ll 1$. This result has important consequences for the rigidity of part of the multiplet structure in the large-$N$ limit, as we clarify in Sec.~\ref{wmm:sec}.}

\subsection{Width of a single multiplet} \label{wsm:sec}
{Let us focus on $\Delta \hat{H}(\alpha,N)=\hat{H}^{(\alpha)}-\hat{H}^{(0)}$,} the difference of the two Hamiltonians {at $\alpha$ and at $\alpha=0$. We choose the basis $\ket{i}$ of eigenstates of $\hat{H}^{(0)}$ such that the matrix elements  $H^{(0)}_{i,j}=\delta_{i,j} E^{(0)}_{S_j}$ with $E^{(0)}_{S}$ denoting the energy of the multiplet with spin $S$ at $\alpha=0$. Then we consider the square root of the quadratic average of the matrix elements of $\Delta \hat{H}(\alpha,N)$, defined in the following way
	\begin{align}\label{medq:eqn}
		\sqrt{\langle\left(H^{(\alpha)}_{i,j} - H^{(0)}_{i,j}\right)^2\rangle} & = \frac{\sqrt{\sum_{i,j}\left(H^{(\alpha)}_{i,j} - H^{(0)}_{i,j}\right)^2}}{\sqrt{\mathcal{N}}}\,.
	\end{align}
$\mathcal{N}$ in the denominator is the number of nonvanishing matrix elements of $\Delta \hat{H}(\alpha,N)$. In order to quantify it we recall that $\Delta \hat{H}(\alpha,N)$ is a sum of terms of the form $\sigma_j^z \sigma_l^z$. Under a global rotation,  $\sigma_j^z \sigma_l^z$ transforms like the sum of a scalar and a tensor, i.e. an object with spin $2$. Thus, by Wigner-Eckart theorem~\cite{Sakurai}, and by the rules of spin addition, we have that, if $\ket{S,i}$ is a state with spin $S$, then  $\sigma_j^z \sigma_l^z\ket{S,i}$ is a superposition of states whose spin is in the set $\{S-2,S-1,S,S+1,S+2\}$. Considering that in each spin-$S$ sector there are $2S+1$ multiplets, and that $\Delta\hat{H}(\alpha,N)$ commutes with the total spin along $z$, we can therefore evaluate $\mathcal{N}$ as 
\begin{equation}\label{N:eqn}
  \mathcal{N}=\sum_{S=0}^{N/2}\sum_{q=\max(-2,S-N/2)}^{\min(2,N/2-S)}(2S+1)g(S)g(S+q)\,.
\end{equation}}

{The numerator in Eq.~\eqref{medq:eqn} is the Hilbert-Schmidt norm of $\Delta \hat{H}(\alpha,N)$, whose symbol is $\left\lVert \Delta {H}(\alpha,N) \right\rVert_{HS}$. As we show in Appendix~\ref{hilbers:sec}, the scaling behavior of this norm is $$\left\lVert \Delta {H}(\alpha,N) \right\rVert_{HS}= \alpha\,\mathcal{K} \sqrt{\dim\mathcal{H}}\,,$$ where $\mathcal{K}>0$ is a numerical factor.}
%
%
	 We emphasize that $\mathcal{K}$ is order $1$ for {the values of} $\alpha<1$ we are considering (see Appendix~\ref{hilbers:sec}). 
$\dim\mathcal{H}{=2^N}$ is the dimension of the full Hilbert space.
	{(Restricting to the fully even subspace will only modify $\dim\mathcal{H}$ and $g(S)$ by a factor $1/N$, leaving Eq.~\eqref{w:eqn} and our conclusions unchanged.)}
	
%
{We assume now that: { (i) the gap{s} separating each multiplet from {the} {neighbouring} ones {are} much larger than the matrix elements coupling it to them; (ii) }
{\em when we restrict to a multiplet}, the spectrum resembles that of a random matrix from the GOE ensemble. We might expect the second assumption to hold {on the one hand due to our results on quantum chaos and on the other hand} since the projection onto a multiplet is an highly non-local operation that will destroy any locality -or sparsity- structure from $H^{(\alpha)}$. { When these assumptions hold}, the eigenvalue spectrum in each multiplet resembles Wigner's semicircle law~\cite{Haake,nothak}, and the multiplet spectral width is given by
	\begin{equation}\label{w:eqn}
		w(N,S) \sim \sqrt{\langle\left(H^{(\alpha)}_{i,j} - H^{(0)}_{i,j}\right)^2\rangle}\, \sqrt{g(S)}=\alpha 2^{N/2}\mathcal{K}\sqrt{\frac{g(S)}{\mathcal{N}}}
	\end{equation}
	with the multiplet-degeneracy $g(S)$ given in Eq.~\eqref{gigi:eqn}, and $\mathcal{N}$ in Eq.~\eqref{N:eqn}. We emphasize that averaging the square matrix elements over all the Hilbert space does not contradict the fact that each multiplet {\em separately} behaves as a random matrix, as long as assumption (i) is valid and there is no mixing between multiplets.} 

{Eq.~\eqref{w:eqn} tells us that our assumption of random-matrix behavior inside a multiplet gives rise to the prediction of a $w(N,S)$ linear in $\alpha$. We can numerically verify that this is exactly what happens for multiplets in the bulk of the spectrum (see Fig.~\ref{plot_multi_w:fig}). So, each multiplet separately behaves as a random matrix and all together give rise to the Wigner-Dyson statistics. Near the edges of the spectrum the behavior is probably different, but states near the spectral edges are a small fraction, vanishing in the limit of large $N$.} 
\begin{figure}
  \begin{center}
   \begin{tabular}{c}
     \includegraphics[width=8cm]{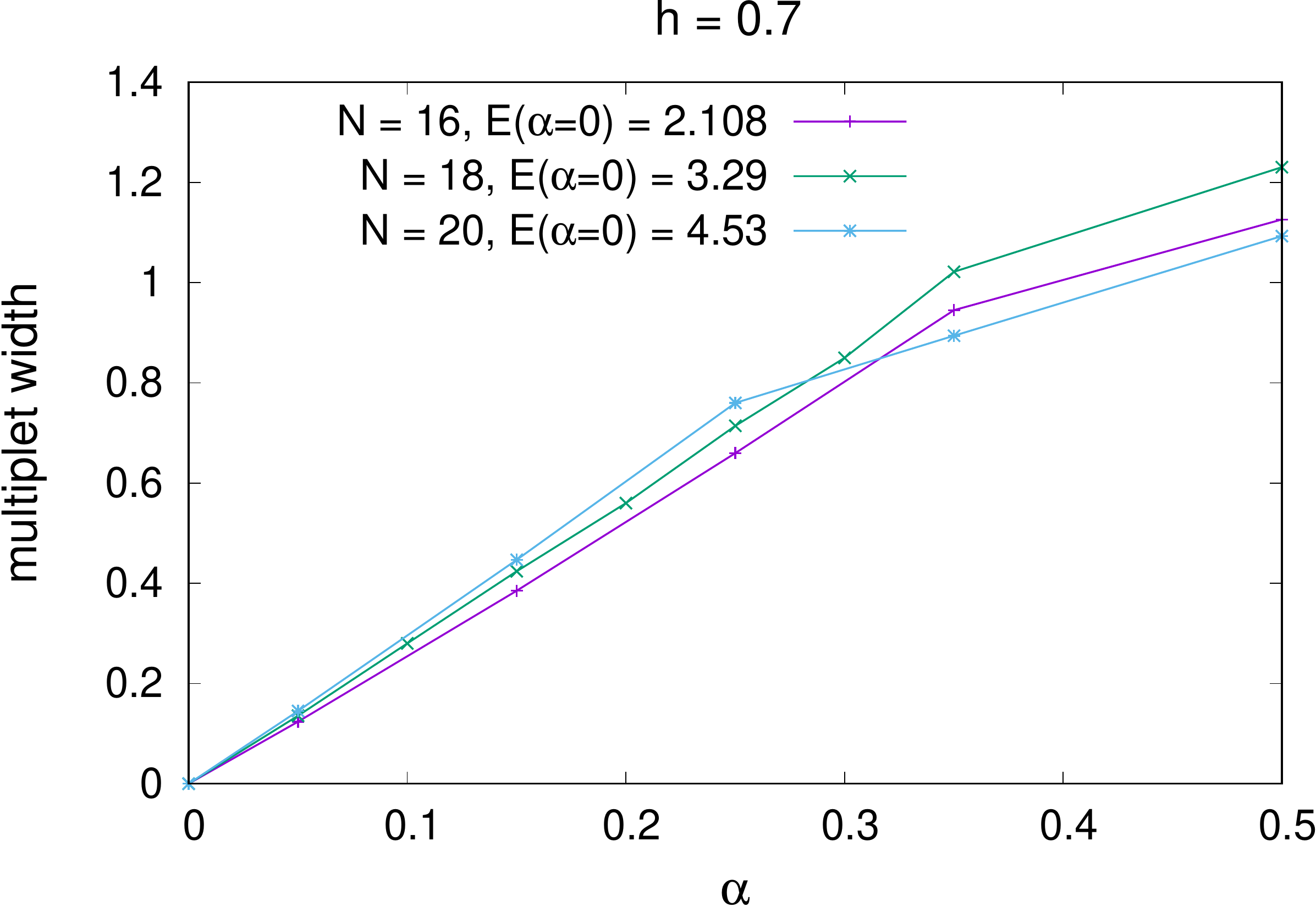}\\
   \end{tabular}
  \end{center}
 \caption{Examples of width of a multiplet versus $\alpha$, for multiplets in the bulk of the spectrum [$E(\alpha=0)$ is the energy of the considered multiplet in the $\alpha=0$ degenerate case]. Notice the linear increase which lasts until the value of $\alpha$ where the considered multiplet starts to overlap with the nearby ones.}
    \label{plot_multi_w:fig}
\end{figure}
\subsection{Total multiplet width and spectral rigidity}\label{wmm:sec}
	In order to {better} understand {the rigidity of} the multiplets {upon increasing system size} $N$, we {now} consider the total multiplet width~\cite{notS} $$W(N)\equiv\sum_{S=0}^{N/2}(2S+1)w(N,S)\,.$$ {We evaluate this quantity using Eqs.~\eqref{w:eqn}~\cite{note_w} and~\eqref{gigi:eqn} and numerically compute the factorials using the Lanczos formula~\cite{recipes}.} We see that $W(N)$ increases linearly in $N$ [see inset of Fig.~\ref{WN:fig}(a)] with a slope obtained from a linear fit $\beta_{W}=0.9$.

{In order to understand if the majority of the multiplets overlaps for large $N$, or {if} there is a significant fraction of them which survives, we need to compare $W(N)$ with the total spectral width $\Delta E(N)\equiv\max_\mu (E_\mu)-\min_\mu (E_\mu)$, which is linear in $N$ with slope $\beta_\Delta\sim 1.1$ (for $h=0.1$) and independent from $\alpha<1$ [see Fig.~\ref{WN:fig}(b)]. So, both $W(N)$ and $\Delta E(N)$ increase linearly in $N$ and their ratio tends to a constant
\begin{equation}
\frac{W(N)}{\Delta E(N)}\;\stackbin[]{N\to\infty}{\longrightarrow}\;\alpha \mathcal{K}\frac{\beta_{W}}{\beta_{\Delta}}\,.
\end{equation}}


{So, when $\alpha<\frac{\beta_\Delta}{\mathcal{K}\beta_W}$, the total multiplet width $W(N)$ is asymptotically smaller than the total spectral width $\Delta E(N)$. In particular, when $\alpha\ll 1$ [more precisely, $\alpha\ll\min(1,\frac{\beta_\Delta}{\mathcal{K}\beta_W})$], we expect that the spectral structure seen in Fig.~\ref{plot_multi:fig} persists for larger system size, with a multiplet structure visible at low energy densities. When $\alpha\ll 1$ we have $W(N)\ll \Delta E(N)$ for large $N$ and we expect that some multiplets persist.} 
\begin{figure}
  \begin{center}
   \begin{tabular}{c}
     \begin{overpic}[width=80mm]{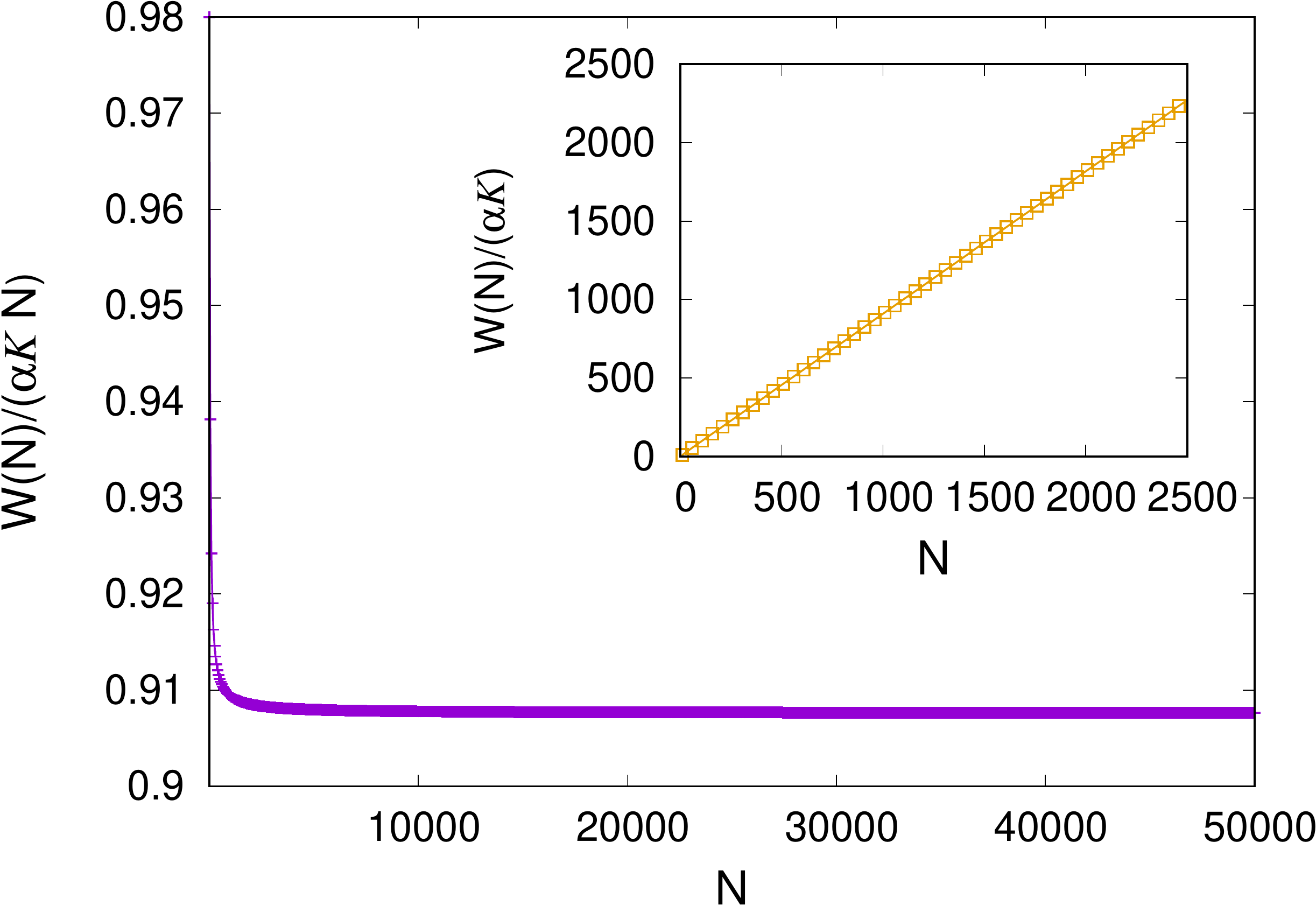}\put(20,53){(a)}\end{overpic}\\
\vspace{-0.1cm}\\
     \begin{overpic}[width=80mm]{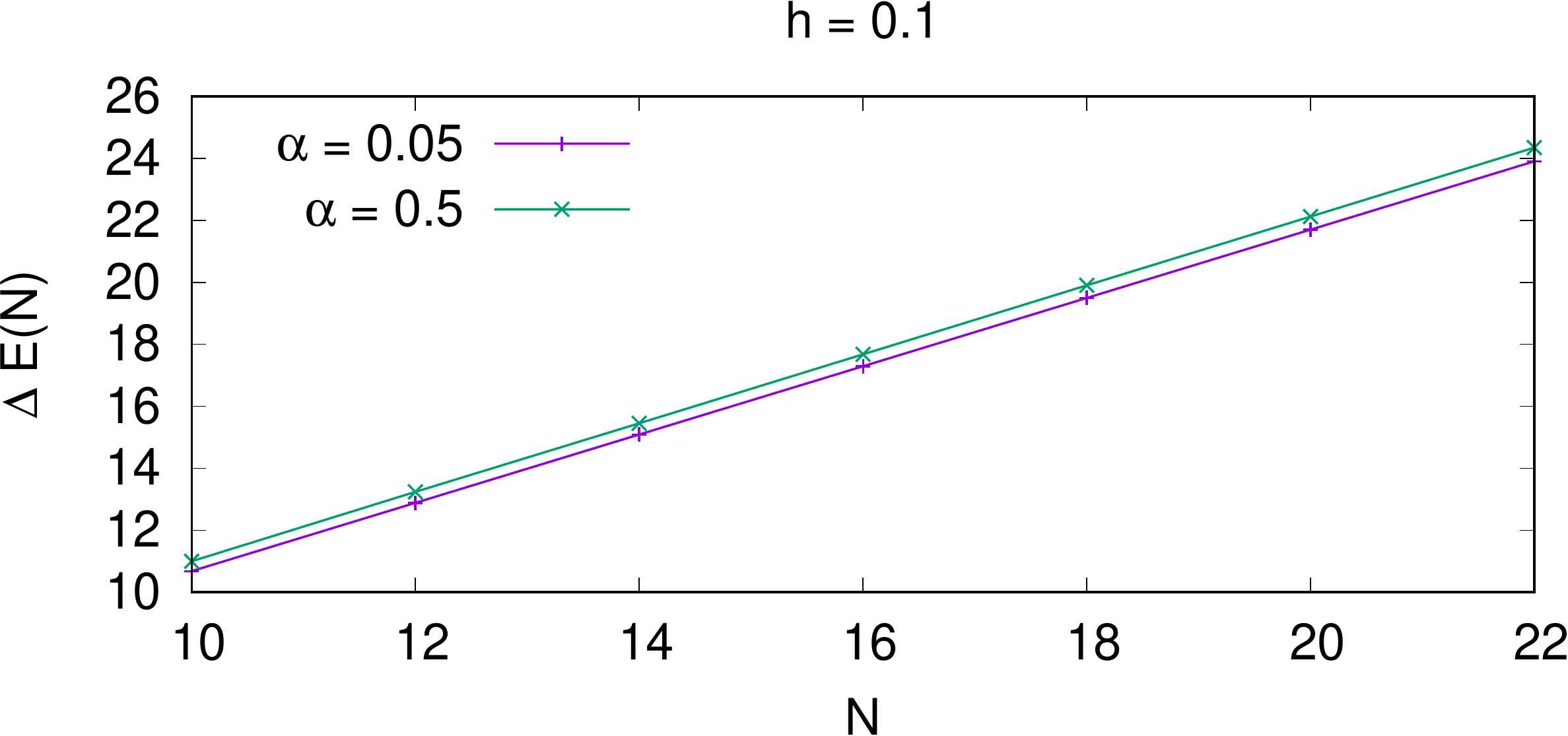}\put(88,13){(b)}\end{overpic}
   \end{tabular}
  \end{center}
 \caption{(Panel a -- main figure) $W(N)/(\alpha\mathcal{K} N)$ versus $N$ for $\alpha<1$. (Inset) $W(N)/(\alpha\mathcal{K})$ versus $N$ for $\alpha<1$. Linear dependence with slope slope  $\beta_W=0.9$. (Panel b) Examples of total spectral width $\Delta E(N)$ versus $N$ for $h=0.1$. $\Delta E(N)$ is defined as the difference between the largest eigenvalue and the smallest eigenvalue of the Hamiltonian restricted to the fully-even Hilbert subspace. The slope $\beta_\Delta\sim 1.1$ comes from a linear fit.}
    \label{WN:fig}
\end{figure}

	{ Looking at Fig.~\ref{plot_multi:fig} (see also Figs.~\ref{DOS:fig} and~\ref{DOS22:fig}), we see that the persisting multiplets lie at low energy densities. The {rigidity of these} multiplets, and the related ensemble inequivalence, are {likely} behind the effective nonergodic behavior and the persistent longitudinal magnetization appearing in low-energy quenches~\cite{Silva,reyhaneh} for $\alpha<2$. }
	
\section{Nonconvex microcanonical entropy and ensemble inequivalence} \label{multiplets:sec}
{The spectrum being organized in multiplets gives rise to a nonconvex microcanonical entropy, with many maxima, one per each multiplet. As we have seen above, for $\alpha\ll 1$, part of the multiplets persists {for very large $N$.} 
A nonconvex microcanonical entropy in this limit gives rise to ensemble inequivalence, as it happens in classical long-range systems~\cite{Mukamel2}.}


{In order to visually show how the presence of multiplets gives rise to a nonconvex microcanonical entropy, let us numerically evaluate the microcanonical entropy $S_{\rm th}(E)$ in a case of finite $N$. To define the entropy, we start from the density of states}
\begin{equation} \label{rho:eqn}
  \rho(E)=\sum_\mu\delta(E-E_\mu)\,.
\end{equation}
We average it over an energy shell (we divide the energy spectrum in $N_{\rm Shell}$ equal energy shells and mark the energy-shell average as $\mean{\cdots}_{\rm Shell}$) and we define $S_{\rm th}(E)=\ln\mean{\rho}_{\rm Shell}(E)$ (for each shell, $E$ is the middle-point energy and we take $k_B=1$). We show our results in Figs.~\ref{DOS:fig} and~\ref{DOS22:fig}.
In Fig.~\ref{DOS:fig}~(a) we plot $S_{\rm th}(E)$ versus the energy density $E/N$ for $\alpha=0.05$, $h=0.1$  and two system sizes. At low and intermediate energy densities, we clearly see the {peaks} corresponding {each} to a multiplet and we do not see {a strong} tendency for them to disappear for increasing system size. We can see something similar for $\alpha=0.25$, $h=0.1$ [Fig.~\ref{DOS:fig}~(b)] where the low and intermediate energy density multiplet structure becomes more evident for increasing system size. So, multiplets strongly affect the dynamics for finite system sizes giving rise to a nonconvex microcanonical entropy. For $\alpha<1$ we clearly see the same nonconvex structure for both $h=0.1$ and $h=0.5$ (Fig.~\ref{DOS22:fig}). {We remark that each peak corresponds to a multiplet, an object with many {levels} giving rise to a smooth random-matrix continuum for $N\to\infty$. So each peak is something physical, very different from the spikes appearing at finite size in the density of states of the short-range Ising model, when a energy shell smaller than the finite-size gaps between the eigenenergies is considered.}

In the plots in Fig.~\ref{DOS:fig} we notice that at the lowest energy densities {we have only few} levels {in the multiplets} and there are significant gaps separating the multiplets. The first two or three multiplets survive even at larger $\alpha$, as we can see in the density-of-states plots of Fig.~\ref{DOS22:fig}, both for $h=0.1$ [panel~(a)] and $h=0.5$ [panel~(b)] where the multiplet structure at intermediate energies is more tight and more fragile to $\alpha>0$. 
\begin{figure}
  \begin{center}
   \begin{tabular}{c}
     \begin{overpic}[width=80mm]{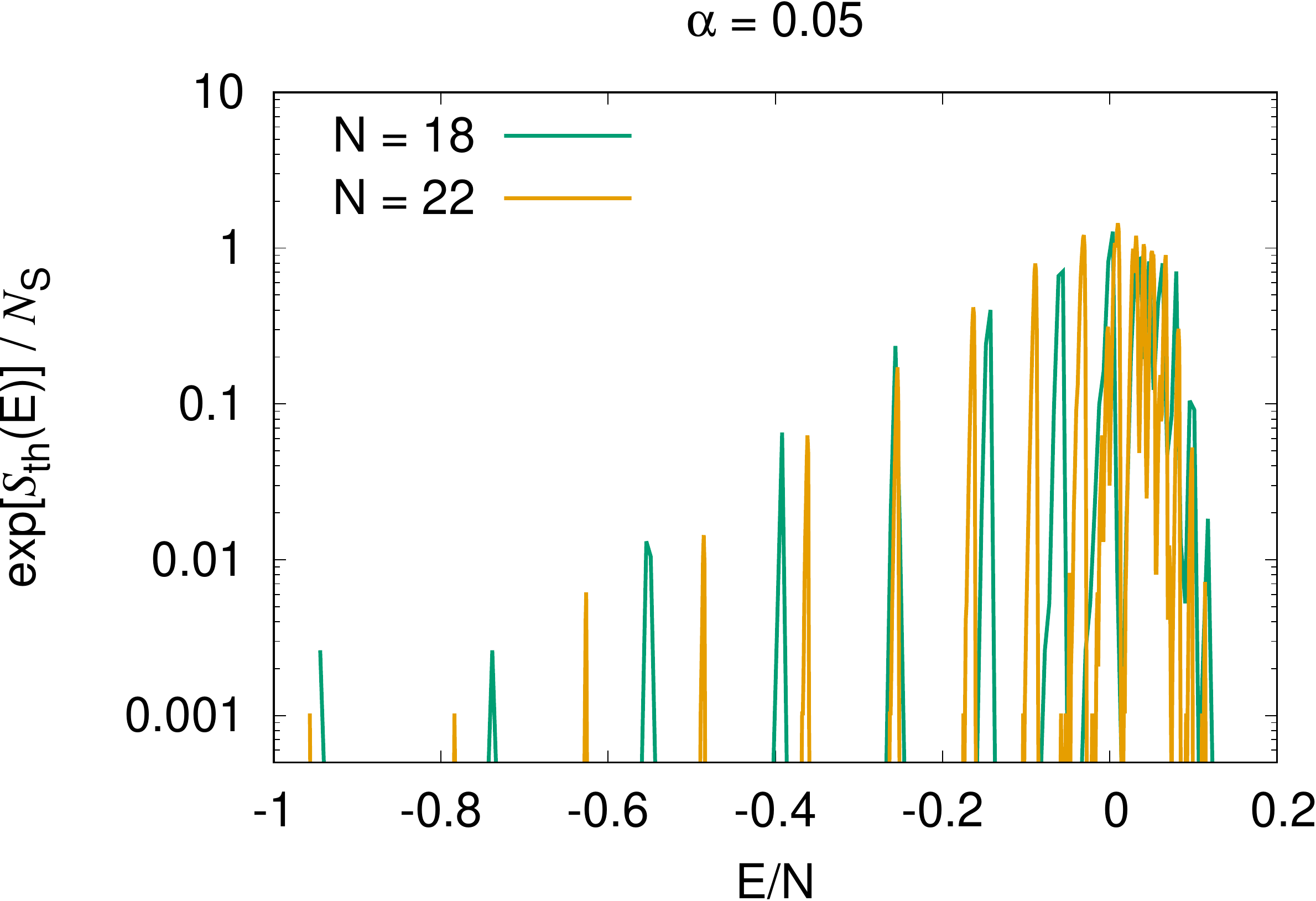}\put(86,55){(a)}\end{overpic}\\
     \vspace{0.1cm}\\
     \begin{overpic}[width=80mm]{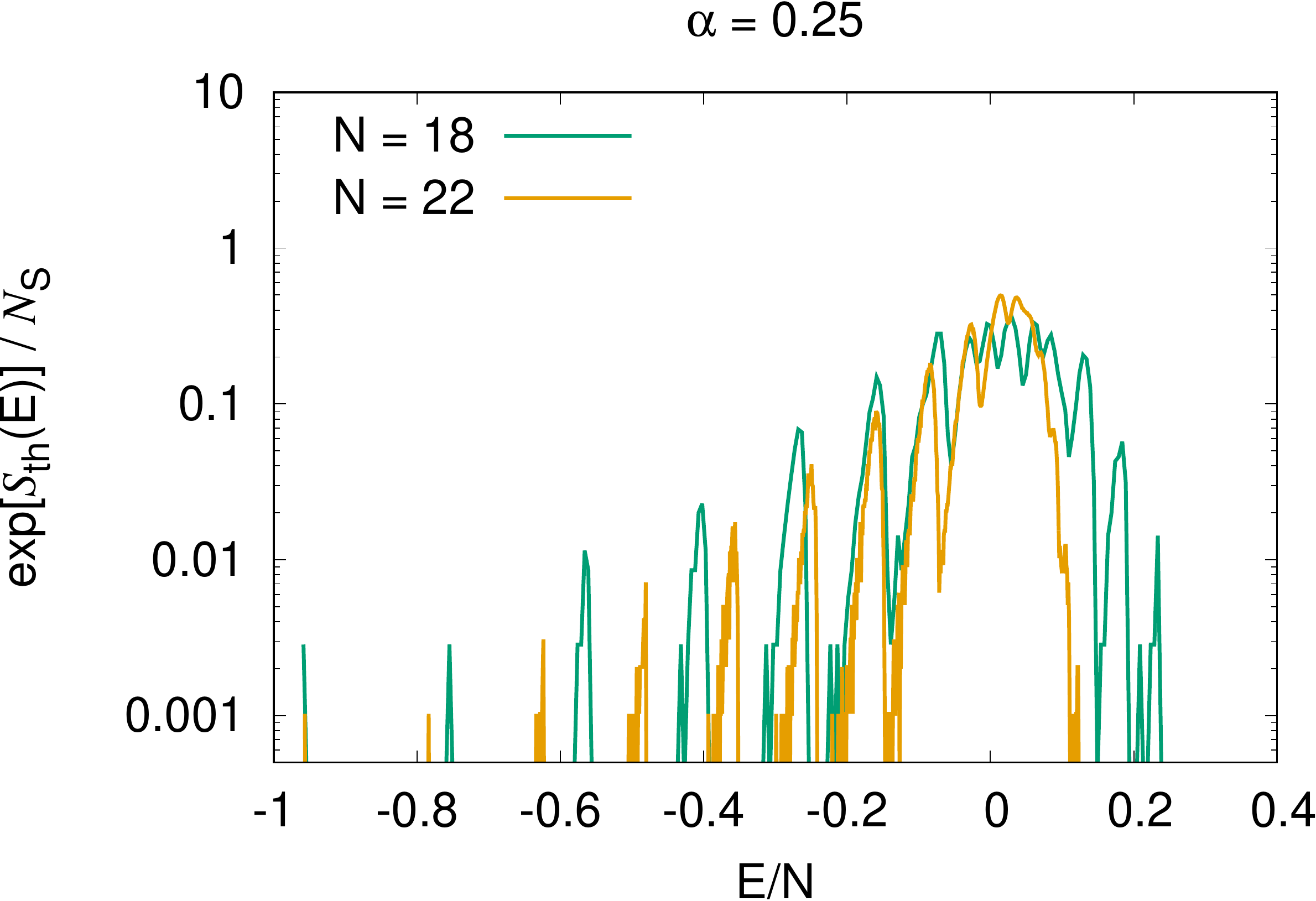}\put(86,55){(b)}\end{overpic}\\
   \end{tabular}
  \end{center}
 \caption{$\exp[S_{\rm th}(E)]/\mathcal{N}_S$ versus $E/N$ for different values of $N$. [Panel~(a)] $\alpha=0.05$, $h=0.1$, $N_{\rm Shell}\geq 200$. [Panel~(b)] $\alpha=0.25$, $h=0.1$,  $N_{\rm Shell}\geq 250$.}
    \label{DOS:fig}
\end{figure}
\begin{figure}
  \begin{center}
   \begin{tabular}{c}
     \begin{overpic}[width=80mm]{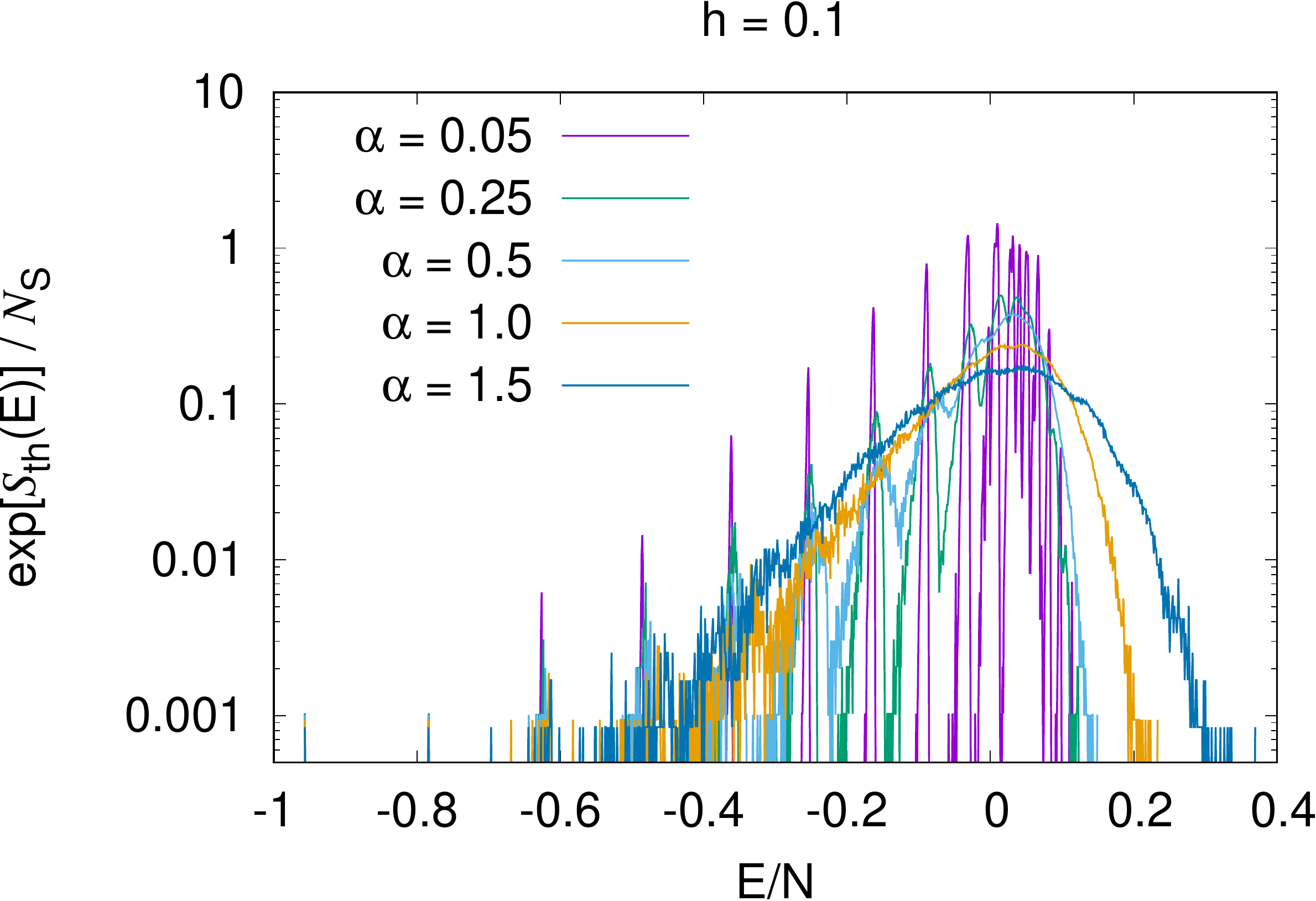}\put(86,55){(a)}\end{overpic}\\
     \vspace{0.1cm}\\
     \begin{overpic}[width=80mm]{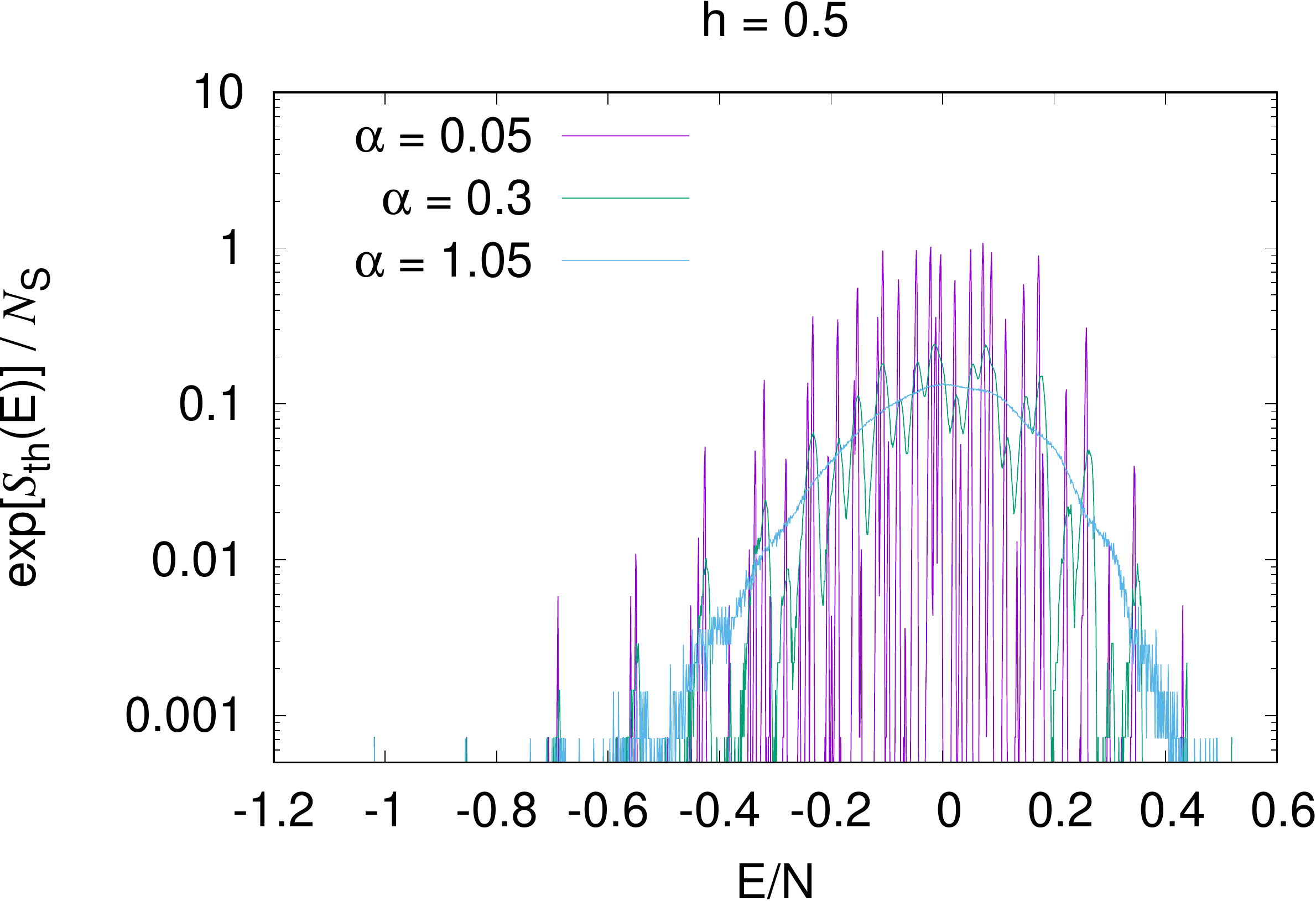}\put(86,55){(b)}\end{overpic}\\
   \end{tabular}
  \end{center}
 \caption{$\exp[S_{\rm th}(E)]/\mathcal{N}_S$ versus $E/N$ for different values of $\alpha$ and $h$. Numerical parameters: $N=22$, $N_{\rm Shell}= 1200$. $h=0.1$ [Panel~(a)] and $h=0.5$ [Panel~(b)].}
    \label{DOS22:fig}
\end{figure}
\section{Spectral pairing and broken symmetry edge} \label{pairing:sec}
It is well known that the long-range quantum Ising chain exhibits a symmetry-breaking transition at nonzero temperature as soon as $\alpha<2$.~\cite{Dutta_PRB2001}
The corresponding microcanonical or even single-eigenstate properties have, however, not been explored extensively, except the notable Ref.~{\cite{fratus2017eigenstate}}.
Here we study the {long-range order of the eigenstates which gives rise to }$\mathbb{Z}_2$ symmetry breaking {in the thermodynamic limit}. In particular, we want to quantify whether for $\alpha\neq 0$ there are states {with long-range order} at finite excitation energy density and to estimate the critical energy density $e^*$ below which the eigenstates break the symmetry in the thermodynamic limit ($e^*$ is called broken symmetry edge~\cite{PhysRevB.86.184303}). The existence of the broken-symmetry edge is well known for the case $\alpha=0$~\cite{PhysRevB.86.184303}, $h<1$, but it is not explored in detail for $\alpha\neq 0$. {We are going to compare this quantity with the corresponding canonical one and show that the two differ from each other for the accessible $\alpha\leq 1.5$ values. 

For {the microcanonical analysis}, we need both the two $\mathbb{Z}_2$ symmetry sectors. Therefore, we restrict to the subspace corresponding to the zero-momentum sector and even only with respect to inversion. We target the single eigenstates and study the energy gaps between nearby states: If there is symmetry breaking {in the thermodynamic limit}, the eigenstates must appear in quasidegenerate doublets, which become degenerate in the thermodynamic limit (the splitting is exponentially small in the system size). {We make use of this property to determine the broken-symmetry edge.} We consider the splitting inside pairs of nearby eigenenergies
%
$\Delta_n^{(1)} = E_{2n}-E_{2n-1}$,
%
($n$ is an integer number labeling the eigenvalues in increasing order) and the gap between nearby pairs, evaluated as the difference of next-nearest neighbor eigenenergies
%
$\Delta_n^{(2)} = E_{2n+1}-E_{2n-1}$.
%
%
If {we are in presence a quasidegenerate doublet ($E_{2n-1}$ and $E_{2n}$ belong to the same doublet)}, $\Delta_n^{(1)}$ should be much smaller than $\Delta_n^{(2)}$ and the ratio $\Delta_n^{(1)}/\Delta_n^{(2)}$ should scale to 0 with the system size. It is convenient to average $\Delta_n^{(1)}$ and $\Delta_n^{(2)}$ on energy shells, in order to reduce fluctuations. {We define the $N_{\rm Shell}$ energy shells as in Sec.~\ref{multiplets:sec} and we consider the ratio
\begin{equation}
  D(E)=\frac{\braket{\Delta_n^{(1)}}_{\rm Shell}(E)}{\braket{\Delta_n^{(2)}}_{\rm Shell}(E)}
\end{equation}
of the averages over the energy shells $\braket{\Delta_n^{(1)}}_{\rm Shell}(E)$ and $\braket{\Delta_n^{(2)}}_{\rm Shell}(E)$.
We term $D(E)$ as the relative splitting and plot it versus $E/N$ for different system sizes in Fig.~\ref{IPR_bin1:fig}. We consider $h=0.1$ and two values of $\alpha$, $\alpha=0.05$ [Fig.~\ref{IPR_bin1:fig}.~(a)] and $\alpha=0.5$ [Fig.~\ref{IPR_bin1:fig}.~(b)]. For the first value of $\alpha$ the spectrum is organized in multiplets for the system sizes we have access to, while for the second it does not. For $\alpha=0.5$ we can see that the curves for different $N$ clearly cross: There is a value of $E/N$ below which $D(E)$ decreases with the system size and above which increases. This is exactly what one would expect for a broken-symmetry edge, and we take this crossing point as an estimate for the broken symmetry edge, with an errorbar given by the mesh in $E$. 

In contrast to the $\alpha=0.5$ case, for $\alpha=0.05$ we do not see any crossing as smooth as this one [Fig.~\ref{IPR_bin1:fig}~(a)]. For this value of $\alpha$ and these system sizes, the dynamics is strongly affected by the above-discussed multiplets. A noisy behavior appears in  Fig.~\ref{IPR_bin1:fig}~(a) and does not allow us to clearly give an estimate for $e^*$. We will estimate the broken symmetry edge only for those values of $\alpha$ and $N$ where we do not see a noisy multiplet structure in the crossing region.

\begin{figure}
  \begin{center}
   \begin{tabular}{c}
      \begin{overpic}[width=80mm]{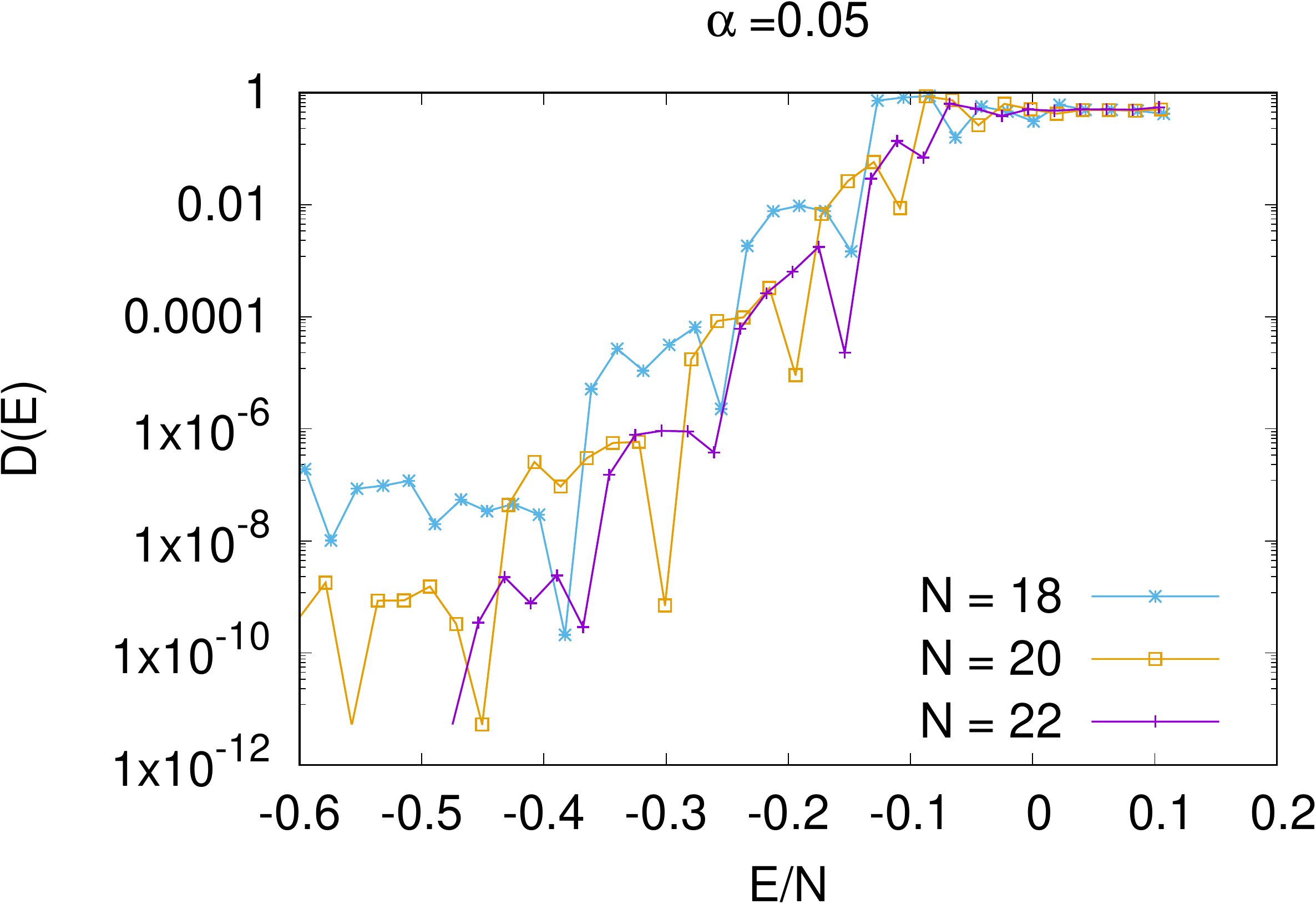}\put(25,53){(a)}\end{overpic}\\
      \\
      \begin{overpic}[width=80mm]{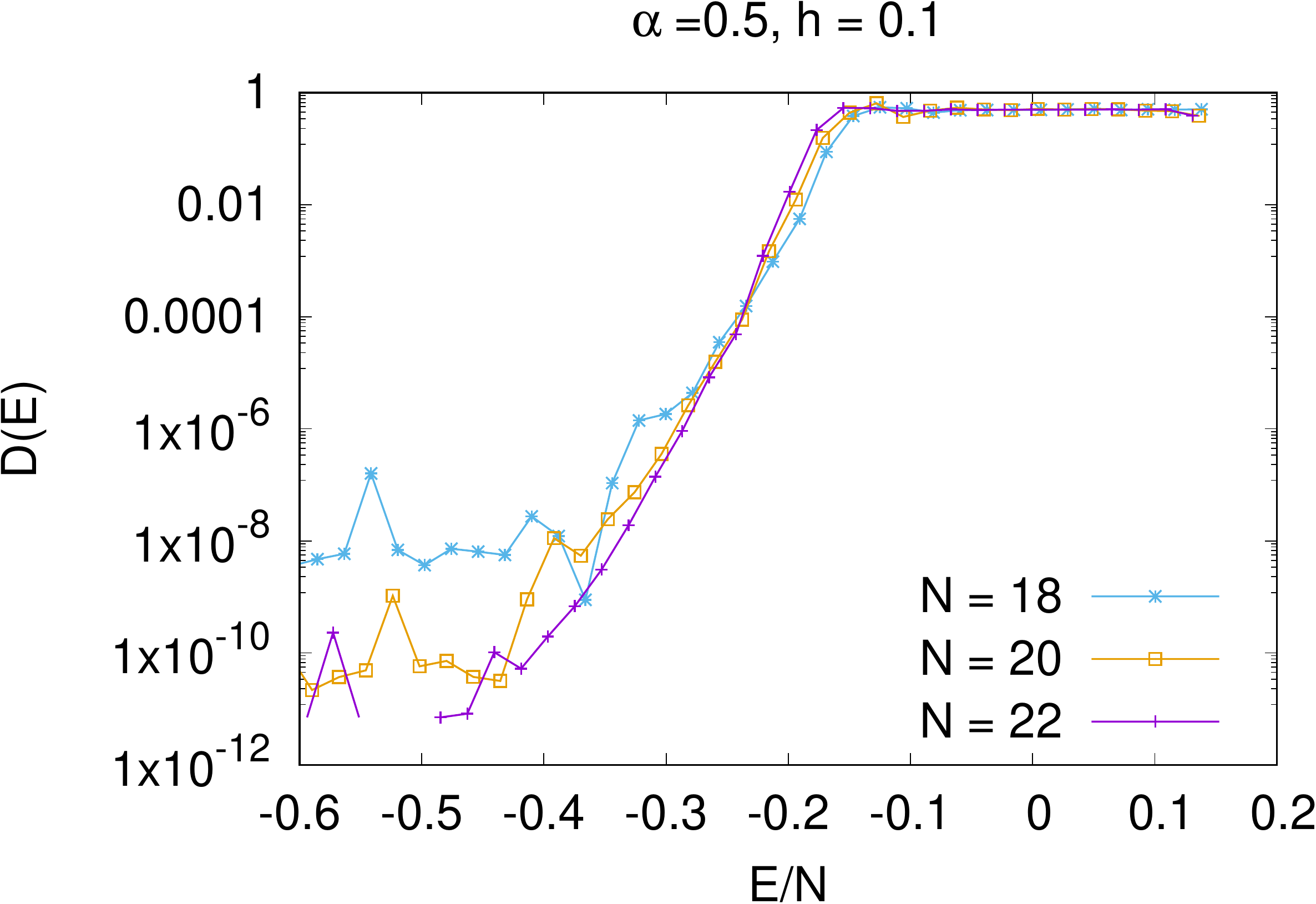}\put(25,53){(b)}\end{overpic}\\
   \end{tabular}
  \end{center}
 \caption{$D(E)$ versus $E/N$ for $\alpha=0.05$ [panel~(a)] and $\alpha=0.5$ [panel~(b)]. $h=0.1$; $N_{\rm Shell}=50$.}
    \label{IPR_bin1:fig}
\end{figure}

{We plot the resulting microcanonical $e^*$ versus $\alpha$ in Fig.~\ref{edge:fig} for $h=0.1$ and $h=0.5$ with the label ``Micro''. We obtain it considering the crossing of the relative-splitting curves for $N=20$ and $N=22$ and for $\alpha=0$ we take the theoretical value $e^*=-h$ found in~\cite{PhysRevB.86.184303}. We can reliably estimate $e^*$ with our method up to $\alpha=1.5$. Above that value larger system sizes are needed.}
\begin{figure*}
 \begin{center}
  \vspace{0.3cm}
  \begin{tabular}{cc}
    \begin{overpic}[width=80mm]{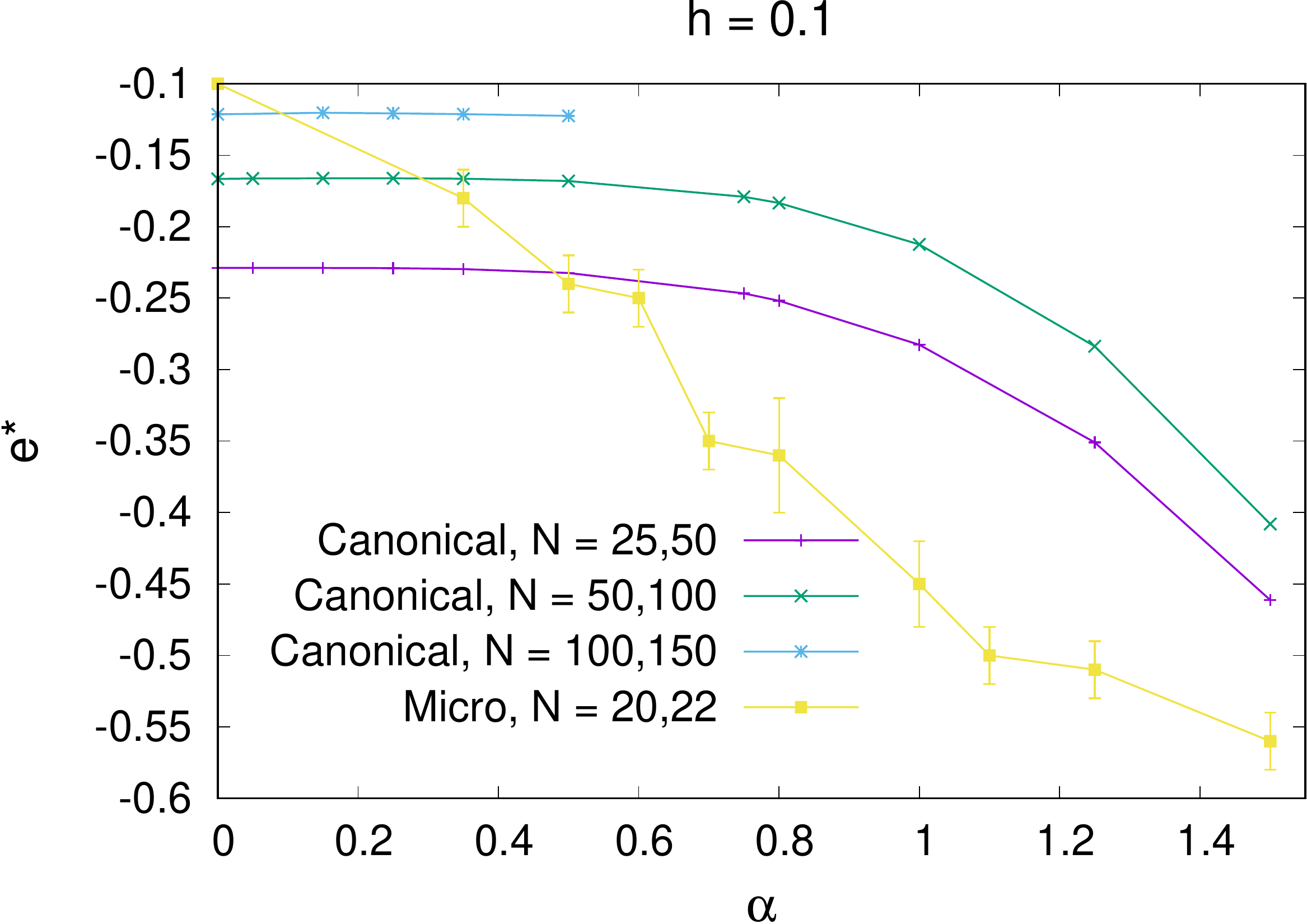}\put(75,53){(a)}\end{overpic}&
    \begin{overpic}[width=80mm]{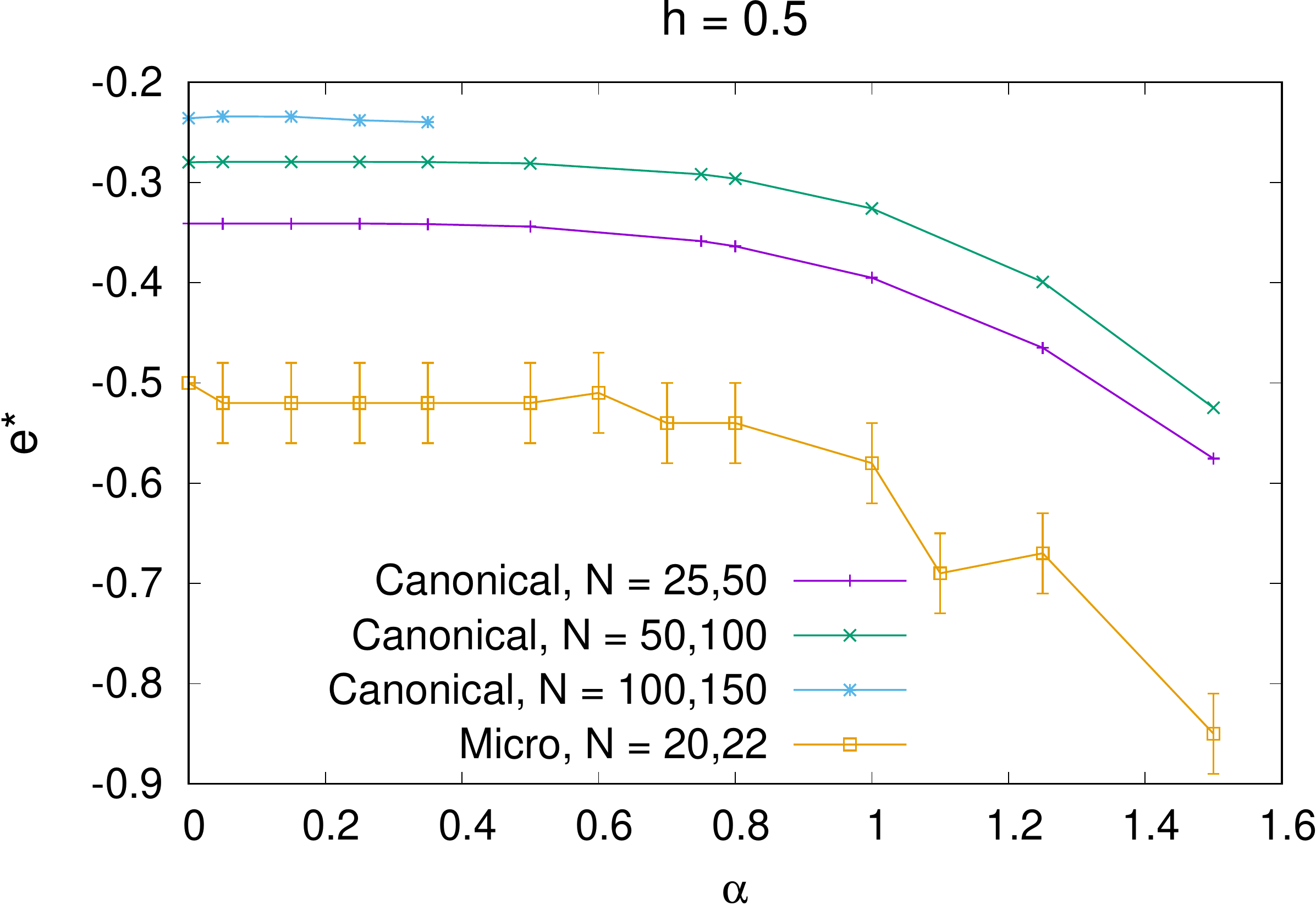}\put(75,53){(b)}\end{overpic}
  \end{tabular}
 \end{center}
 \caption{ Microcanonical $e^*$ versus $\alpha$ for two values of $h<1$ (label ``Micro'', from the crossing of $D(E)$ versus $E/N$ curves) compared with the corresponding canonical value (label ``Canonical'', obtained from the crossing of the Binder-cumulant curves). In the captions the values of $N$ of the two crossing curves are specified. In the canonical case, the step of the imaginary time evolution is everywhere $\tau=10^{-3}$  but in the curve ``Canonical, N = 100,150'' in panel (a) where $\tau=2\cdot 10^{-3}$.}
    \label{edge:fig}
\end{figure*}

{We compare it with the canonical broken-symmetry edge labeled as ``Canonical''  in Fig.~\ref{edge:fig}. The latter is evaluated considering the Binder cumulant, a measure of $\mathbb{Z}_2$ symmetry breaking particularly effective in the canonical ensemble~\cite{Binder}. Defining $\hat{S}_z^q\equiv\left(\sum_j\hat{\sigma}_j^z\right)^q$, the Binder cumulant is given by
%
  $B=1-\frac{\braket{\hat{S}_z^4}_{\rm th}}{3\braket{\hat{S}_z^2}_{\rm th}^2}$,
%
where $\braket{\cdots}_{\rm th}$ is the thermal canonical expectation. Varying the temperature, both $B$ and the corresponding energy density $e=\braket{\hat{H}}_{\rm th}/N$ vary. We plot $B$ versus $e$ for a set of parameters and two different values of $N$ in Fig.~\ref{bincan:fig}. The canonical symmetry breaking threshold is estimated as the crossing between these two curves, in a way similar to what done in~\cite{reyhaneh}. Here the thermal canonical expectations $\braket{\cdots}_{\rm th}$ are obtained by evolving in imaginary time a purified infinite-temperature state~\cite{PhysRevB.94.115157, PAECKEL2019167998}. The imaginary-time evolution is performed through the TDVP algorithm~\cite{PhysRevLett.107.070601,PhysRevB.94.165116}.} 
\begin{figure}
 \begin{center}
  \begin{tabular}{c}
    \includegraphics[width=8cm]{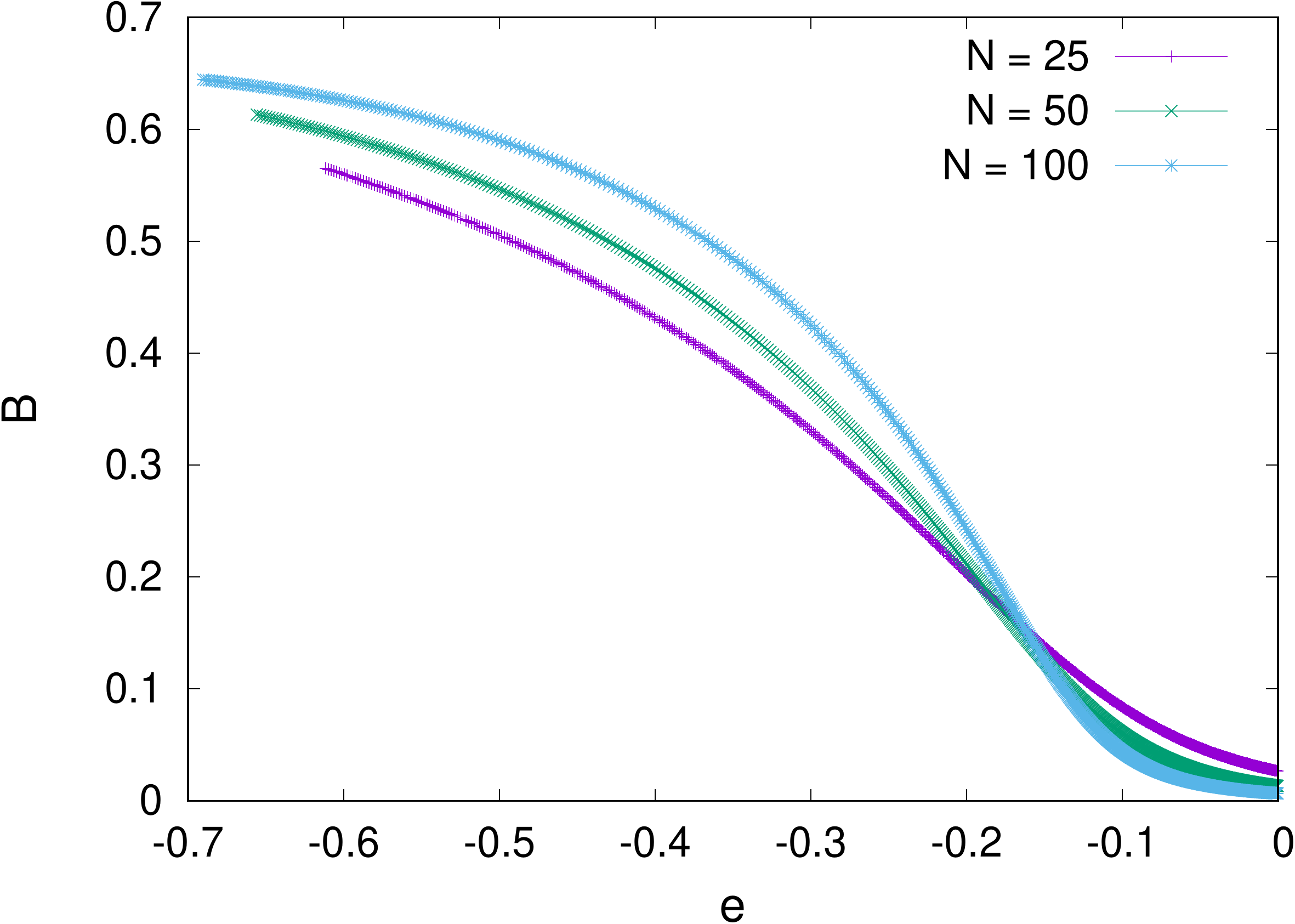}
  \end{tabular}
 \end{center}
 \caption{Canonical Binder cumulant versus the corresponding thermal energy density. Results obtained with TDVP (see main text), time step of the imaginary evolution $\tau=10^{-3}$.}
    \label{bincan:fig}
\end{figure}

{The canonical $e^*$ versus $\alpha$ (Fig.~\ref{edge:fig}) shows a strong dependence on $N$ and so that the canonical $e^*$ increases if we take the crossing of curves for larger $N$: The difference with the microcanonical value increases. This fact suggests ensemble inequivalence, but finite-size effects are too strong for making a precise statement.}

{Moreover, considering that the ground-state is at $e_{\rm GS}\simeq -1$), Fig.~\ref{IPR_bin1:fig} gives us the nontrivial conclusion that for $\alpha\leq 1.5$ the system shows $\mathbb{Z}_2$ symmetry breaking at finite excitation energy densities. So, there is a finite fraction of {the energy-spectrum width where the eigenstates show long-range order}, similarly to the $\alpha=0$ and the disordered case. This is in agreement with the findings of~\cite{Silva,reyhaneh}, where the long-time dynamics supports long-time magnetization in the range $\alpha\leq 1.5$ and beyond. }
\section{ETH properties} \label{nearco:sec}
%
{After having studied in detail spectral properties, we now take a step further and aim to}
{study eigenstate-thermalization properties. {For concreteness,} we consider the longitudinal nearest-neighbour correlation operator
\begin{equation}
  \hat{\mathcal{G}}=\frac{1}{N}\sum_{j=1}^N\hat{\sigma}_j^z\hat{\sigma}_{j+1}^z\,,
\end{equation}
as a representative for local observables. We focus on the properties of the eigenstate expectation {values}   $\mathcal{G}_\mu\equiv\braket{\varphi_\mu|\hat{\mathcal{G}}|\varphi_\mu}$. {We expect that the same behaviour occurs for any local observable. As we show in Appendix~\ref{entropies:sec} also the entanglement entropy (involving half of the system size) shows a similar behaviour.}
%

%

%
}

{We consider the scatter plots of $\mathcal{G}_\mu$ versus $E_\mu$ in Fig.~\ref{plots_oper:fig}. Most importantly, these expectation values as a function of energy don't always exhibit a smooth dependence with small fluctuations, as expected in a  system obeying ETH~\cite{Rigol_Nat} even though the level spacing ratio Eq.~\eqref{rorro:eqn} is close to Wigner-Dyson. The finite-size effects are too strong, mainly related to the spectrum being organized in multiplets for $\alpha<1$, and no quantitative extrapolation to larger size is possible. Nevertheless we see a lack of correspondence between quantum chaos and ETH, in contrast with short-range interacting systems. 

{The most noteworthy case is $\alpha=0.05$ [Fig.~\ref{plots_oper:fig}~(a) and~(b)] where we see many almost vertical lines, as many as the multiplets. Each of these lines is a continuous curve, {as if} ETH {was to hold} just within a multiplet but not across them. As we have argued in Sec.~\ref{randma:sec}, when $N$ is increased, part of the multiplets {should} survive, {and then this behaviour should persist}. What we see in Fig.~\ref{plots_oper:fig}~(a) and~(b) is {nevertheless} strongly affected by finite size effects. 

{Another interesting case is provided by $\alpha=0.5$ [Fig.~\ref{plots_oper:fig}~(c) and~(d)]. For $h=0.1$ [panel~(c)] we can see a qualitatively different behavior at large and small energy. In the center of the spectrum we observe a quite smooth curve with some {small} fluctuations, which appears as a prototypical example of a system obeying ETH. 
{Overall, for these small system sizes, this doesn't seem to follow the predictions by ETH.}
\begin{figure*}
  \begin{center}
   \begin{tabular}{cc}
     \hspace{-1cm}\begin{overpic}[width=80mm]{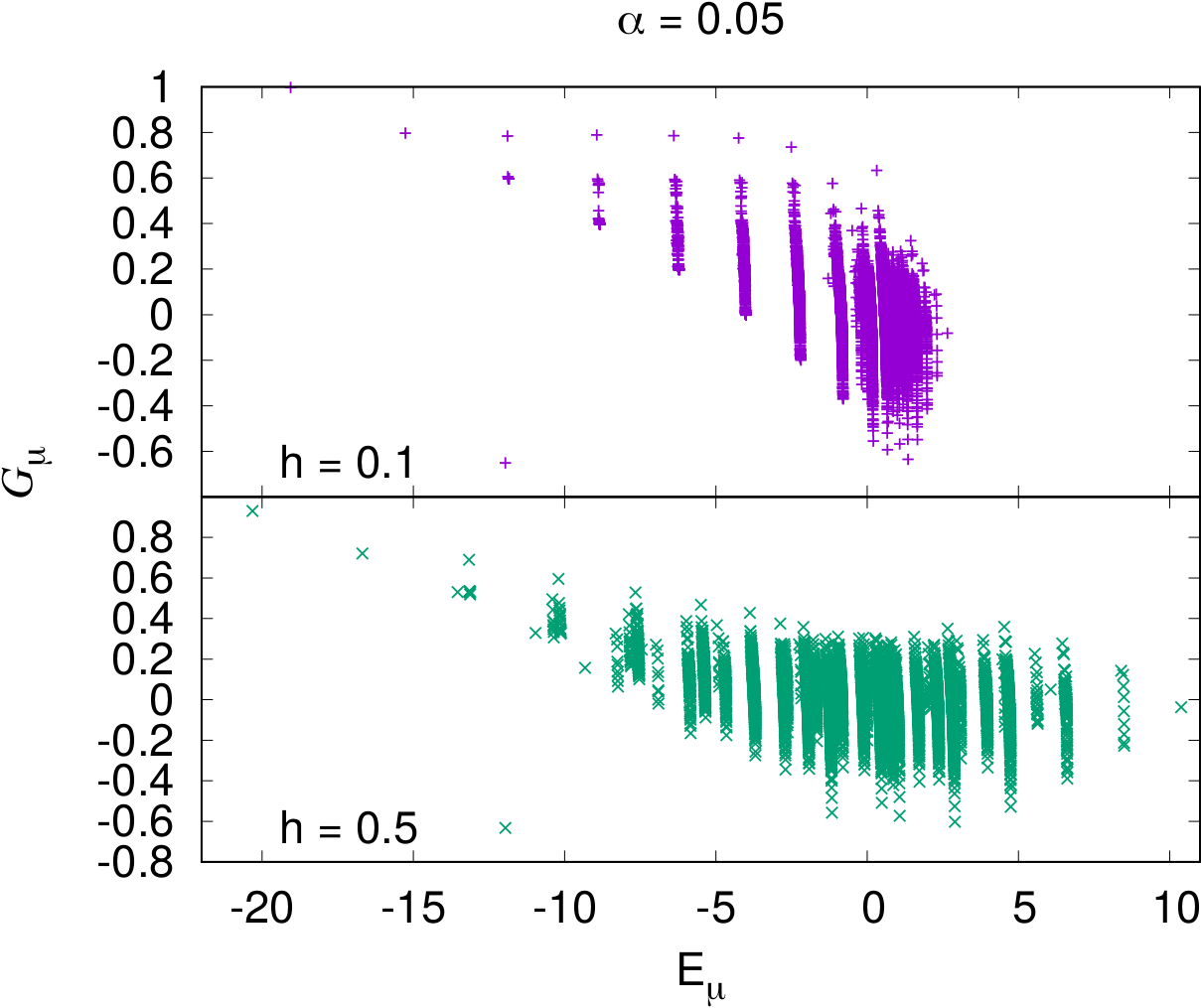}\put(90,50){(a)}\put(90,35){(b)}\end{overpic} &
     \hspace{0.5cm}\begin{overpic}[width=80mm]{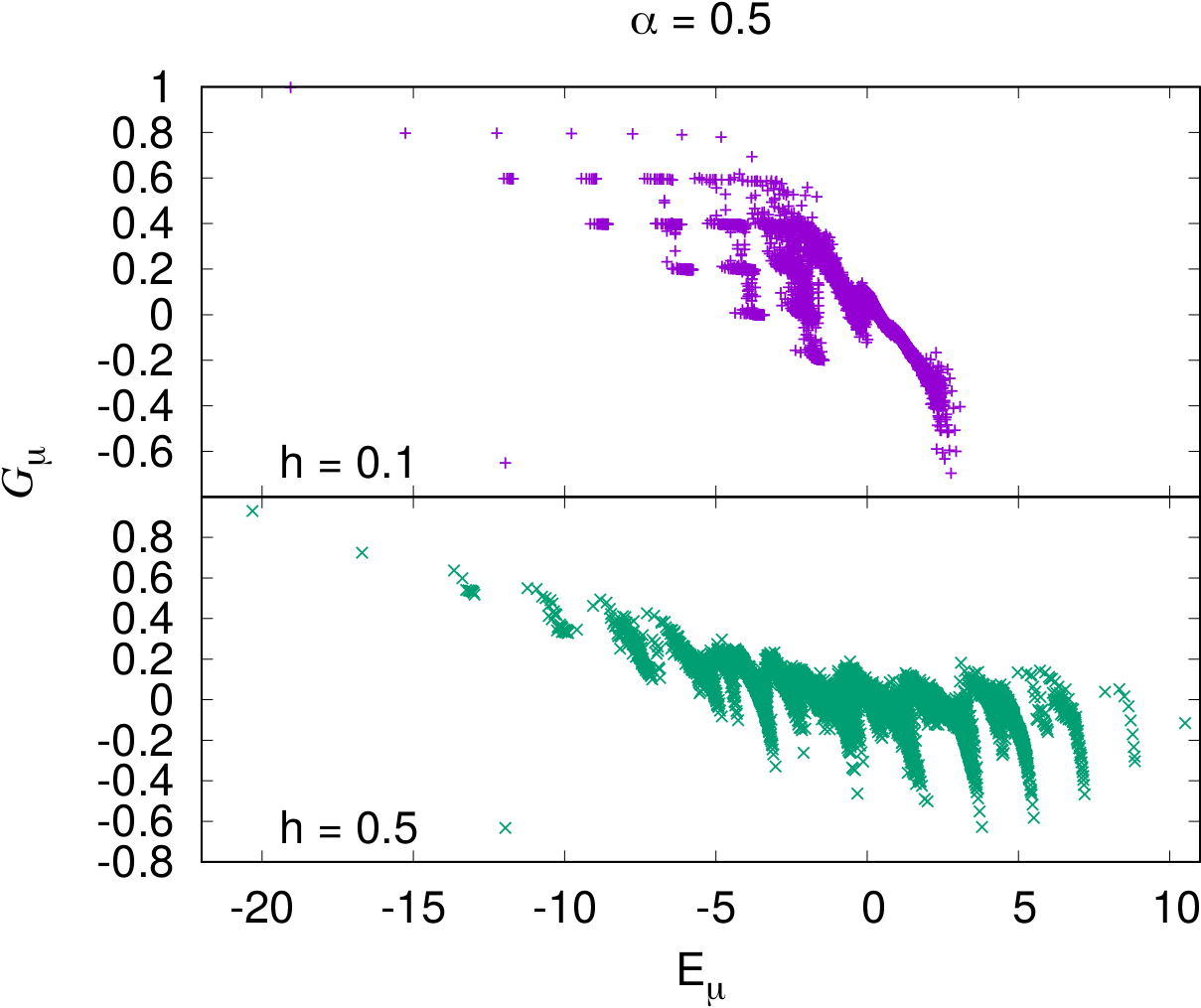}\put(90,50){(c)}\put(90,35){(d)}\end{overpic}\\
     \\
     \hspace{-1cm}\begin{overpic}[width=80mm]{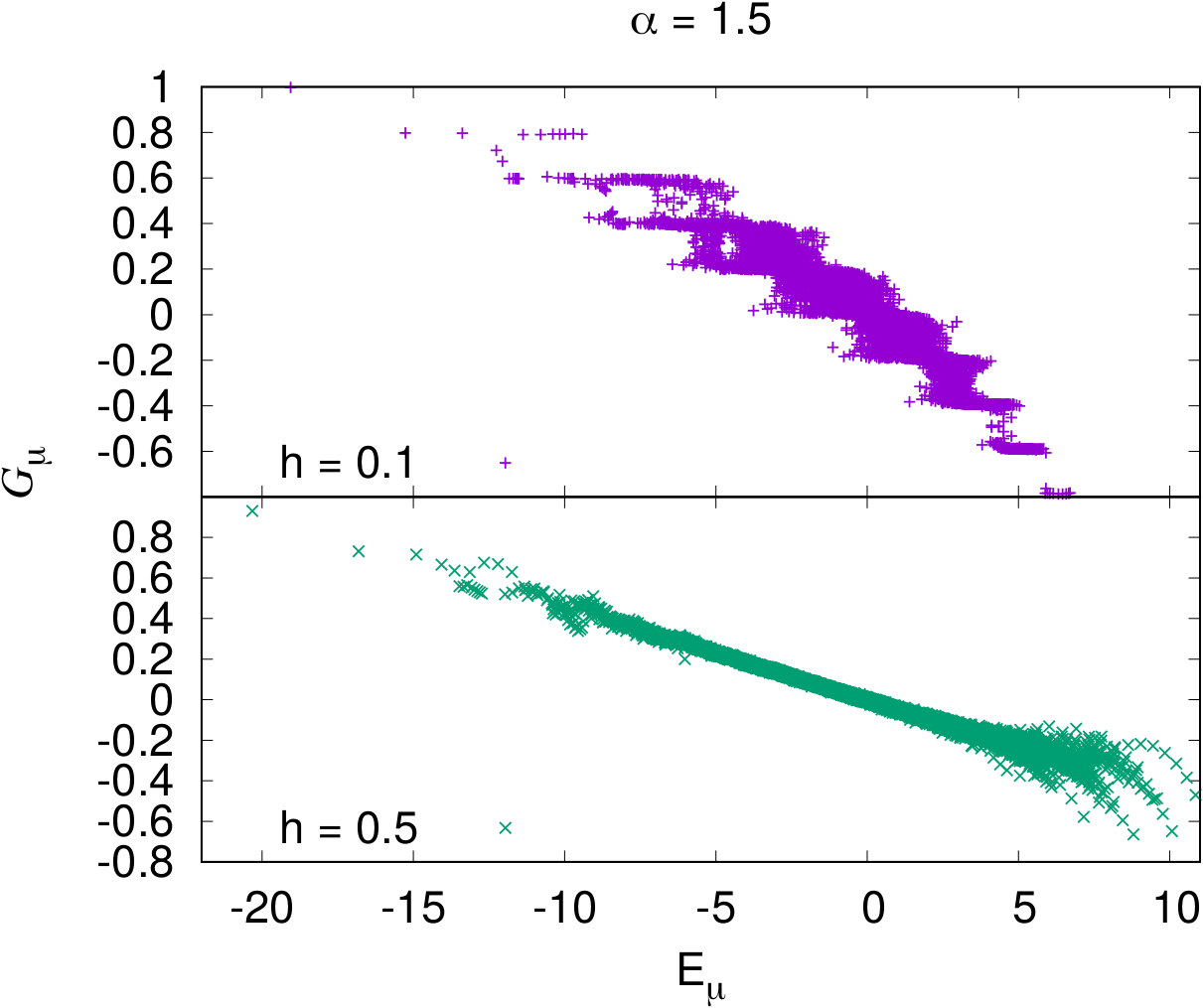}\put(90,50){(e)}\put(90,35){(f)}\end{overpic}&
     \hspace{0.5cm}\begin{overpic}[width=80mm]{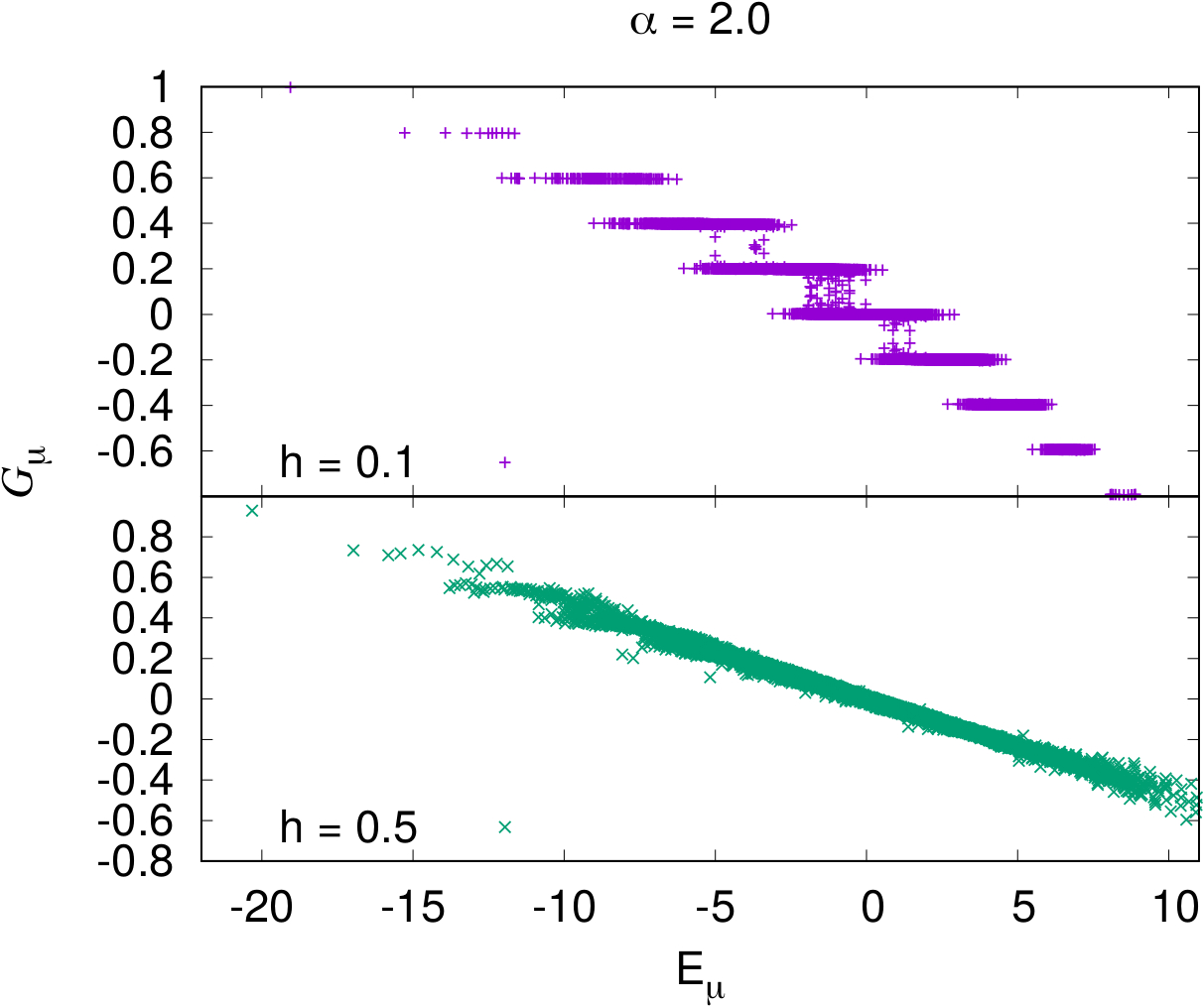}\put(90,50){(g)}\put(90,35){(h)}\end{overpic}\\
   \end{tabular}
  \end{center}
 \caption{Scatter plots of $\mathcal{G}_\mu$ versus $E_\mu$ for different values of the parameters. We consider $N=20$.}
    \label{plots_oper:fig}
\end{figure*}

{For larger $\alpha$ [$\alpha=1.5$ in Fig.~\ref{plots_oper:fig}~(e),~(f) and $\alpha=2$ in Fig.~\ref{plots_oper:fig}~(g),~(h)] we see a fully developed ETH behavior for $h=0.5$: very smooth curves with noise at the edges of the spectrum [panels~(f) and~(h)]. On the opposite, for $h=0.1$ [panels~(e) and~(g)], the situation is not at all ETH, in close correspondence with the average level spacing ratio being different from Wigner-Dyson [Fig.~\ref{correl1:fig}~(a)]. In particular, the case $\alpha=2$ is very regular-like {with} some {scattered points} {between the horizontal {lines}} suggesting a stronger mixing at larger system sizes. 
%


%
\section{Conclusion}\label{conc:sec}
In conclusion we have considered the long-range Ising model with power-law interactions and used exact diagonalization to study the relation between quantum chaos, eigenstate thermalization and convexity of the microcanonical entropy. For small $\alpha$ we have remarkably found that the level spacing distribution is Wigner Dyson but this does not reflect a full-random-matrix behaving Hamiltonian. 

{The reason comes from the strong effect of the $\alpha=0$ integrable point, where the Hilbert space decomposes into many identical subspaces with the same energy levels, {due to the full permutation symmetry}. Even an infinitesimal $\alpha> 0$ mixes the degenerate levels belonging to different subspaces; the resulting spectrum is organized in multiplets and we argue that multiplets in the bulk of the spectrum separately behave as a random matrices, {with a negligible role of the spectral edges}.}

Due to the strong effect of multiplets, this Wigner-Dyson spectral statistics appears in association with anomalous thermalization properties. The random-matrix behavior of the multiplets {suggests} that part of the multiplets persists at large $N$ and $\alpha<1$.  This holds in particular in the  $\alpha\ll 1$ limit.} {So, also at large $N$ there are multiplets, and they give rise to a nonconvex {microcanonical} entropy {as a function of} energy, imply{ing} ensemble inequivalence~\cite{Mukamel2}.} From the numerics, we expect that the multiplets persisting at large $N$ lie at low energy densities; they are probably involved in the persistent magnetization, which has been observed in the low-energy dynamics of this model~\cite{Silva,reyhaneh}.

We {further analyse} the eigenstate thermalization properties and we see that at small $\alpha$ the local observable eigenstate expectation {value}s and the corresponding half-system entanglement entropies do not organize in{to} smooth curves {as a function of the} energy, as one should naively expect from {quantum chaotic behavior in} the Wigner-Dyson level spacing statistics. 
In contrast to short-range interacting systems the spectrum is organized in multiplets and there is no simple ETH behavior. Quantitative probes (see Appendix~\ref{entropies:sec}) suggest that the curves become smoother for increasing system sizes and we cannot tell if this is due to the ETH being obeyed better and better inside the multiples or to the fact that the multiplets at large energy densities tend to merge.

We remark that our exact diagonalization results show a persisting nonergodic behavior for $h=0.1$ and $\alpha$ around the value $\alpha \approx 2$. This is a suggestive result because there are other long-range models with $\alpha=2$ which are integrable, but the system sizes we have access to do not allow to state if this effect persists in the thermodynamic limit. Nevertheless, a nonchaotic behavior for $N=22$ is already remarkable and might suggest {at least} the proximity of an integrable point. In all the other cases we see an ergodic behavior.

{Perspectives of future work will focus on the connection between the dynamical phase transition in $\alpha$ undergone by this model~\cite{Silva,reyhaneh,2017Halimeh} and the corresponding low-energy confinement-deconfinement transition~\cite{PhysRevLett.122.150601}. Another direction of research will be to study the relation between quantum chaos in sectors of the Hilbert space and ensemble inequivalence in models with Hilbert space fragmentation~\cite{Hahn2021}.}
\begin{acknowledgements}
We thank M.~Dalmonte, R,~Khasseh, S.~Pappalardi, F.~M.~Surace and especially R.~Fazio for fruitful discussions. We gratefully acknowledge M.~Kastner and G.~Piccitto for insightful comments on the manuscript. A.~R. warmly thanks D.~Rossini and A.~Tomadin for the access to the late GOLDRAKE cluster where part of the numerical work for this project was performed.
This project has received funding from the European Research Council (ERC) under the European Union's Horizon 2020 research and innovation programme (grant agreement No. 853443), and M. H. further acknowledges support by the Deutsche Forschungsgemeinschaft (DFG) via the Gottfried Wilhelm Leibniz Prize program.

\end{acknowledgements}
\appendix
%
\section{Eigenstate half-system entanglement entropies} \label{entropies:sec}
ETH properties of eigenstates can be explored also by means of the entanglement entropy.
This is not a local object because it involves correlations extending up to a distance $N/2$, but eigenstate thermalization has been proved valid for subsystems up to this size~\cite{PhysRevB.93.134201}. Considering an eigenstate $\ket{\varphi_\mu}$, and {decomposing} the system in two parts $A$ and $B$ in physical real space, we define
\begin{equation} \label{sent:eqn}
  S_{A}^{(\mu)} = -\Tr[\hat{\rho}_A\log\hat{\rho}_A]\quad{\rm with}\quad \hat{\rho}_A=\Tr_B[\ket{\varphi_\mu}\bra{\varphi_\mu}]\,.
\end{equation}
{Specifically,} we focus on the half-system entanglement entropy $S_{N/2}^{(\mu)}$ {taking} each bipartition made {up of} $N/2$ consecutive spins. In case of eigenstate thermalization, $S_{N/2}^{(\mu)}$ are equal to their microcanonical value at energy $E_\mu$, up to relative fluctuations decreasing with the system size. (The microcanonical {value} of $S_{N/2}^{(\mu)}$ is the microcanonical entropy {for half of the system}.)

In Fig.~\ref{correl2:fig} we show the scatter plots of the entanglement entropy $S_{N/2}^{(\mu)}$ [defined in Eq.~\eqref{sent:eqn}] versus the corresponding eigenstate energy  $E_\mu$. ETH is strictly related to these curves looking ``smooth'', as appropriate for microcanonical entropy~\cite{PhysRevB.93.134201}. Let us first discuss this point qualitatively. We consider a small value of $\alpha$, $\alpha=0.05$ [panels~(a),~(c)].  The $S_{N/2}^{(\mu)}$ versus $E_\mu$ look like smooth curves, as in the ETH case, only if we restrict inside the multiplets. This result fits with the average level spacing ratio being Wigner-Dyson for these small values of $\alpha$ (Sec.~\ref{spectrum:sec}) and each multiplet behaving separately as a random matrix (Sec.~\ref{randma:sec}). The nonconvex entanglement entropy of these curves corresponds to a nonconvex microcanonical entropy and to ensemble inequivalence (see Sec.~\ref{multiplets:sec}).
\begin{figure*}
  \begin{center}
   \begin{tabular}{cc}
      \hspace{-1cm}\begin{overpic}[width=80mm]{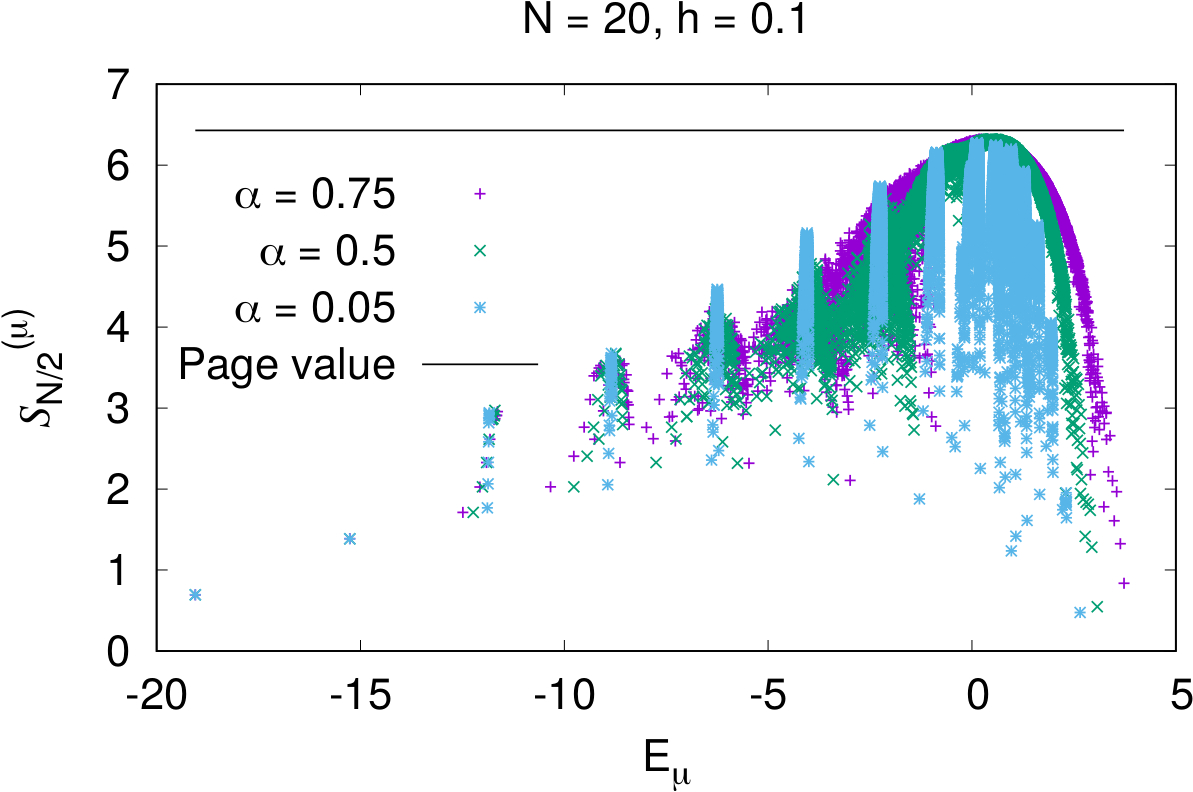}\put(90,50){(a)}\end{overpic}&
      \begin{overpic}[width=80mm]{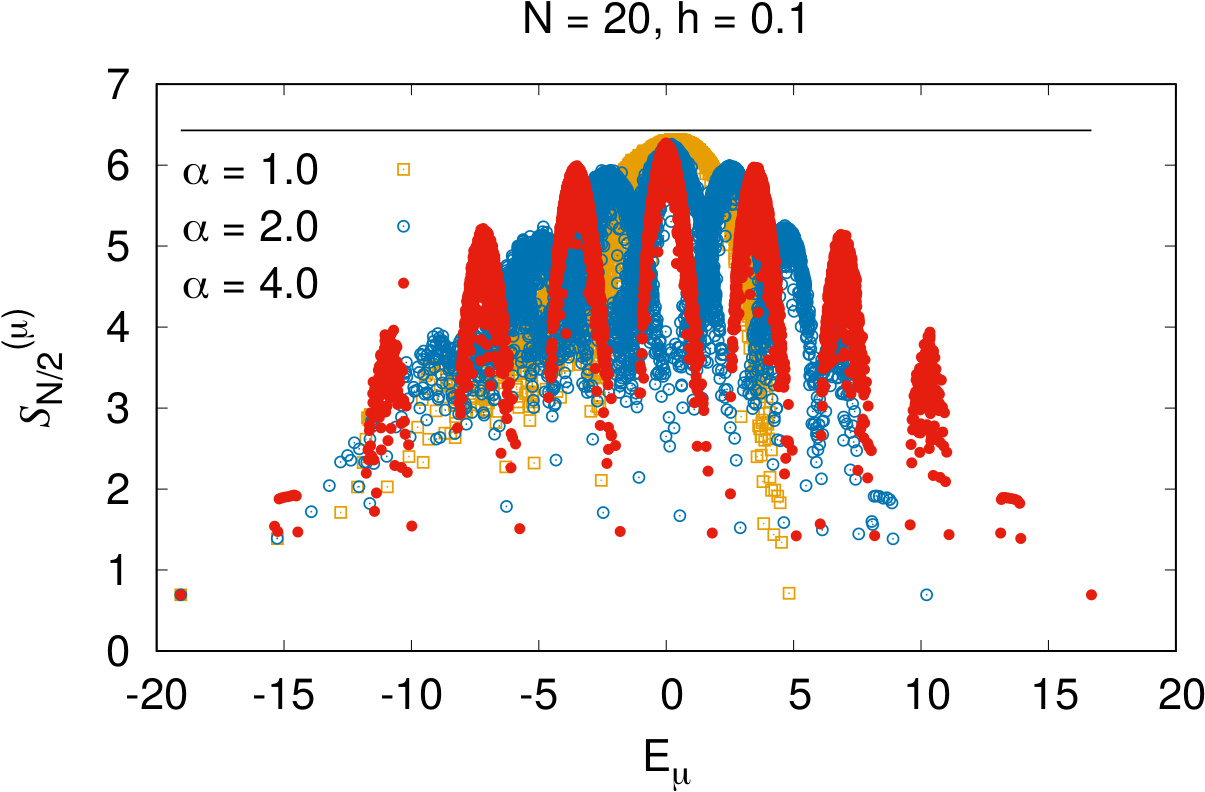}\put(90,50){(b)}\end{overpic}\\
      \\
      \hspace{-1cm}\begin{overpic}[width=80mm]{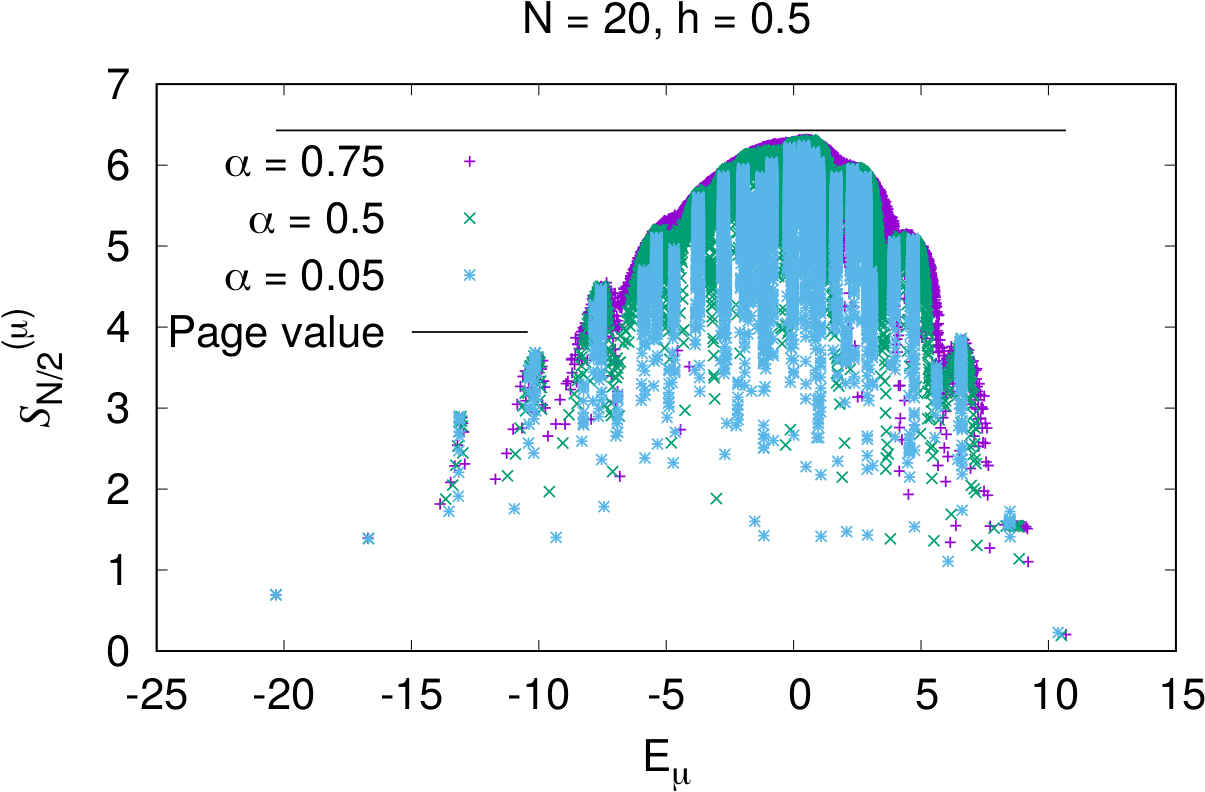}\put(90,50){(c)}\end{overpic}&
      \begin{overpic}[width=80mm]{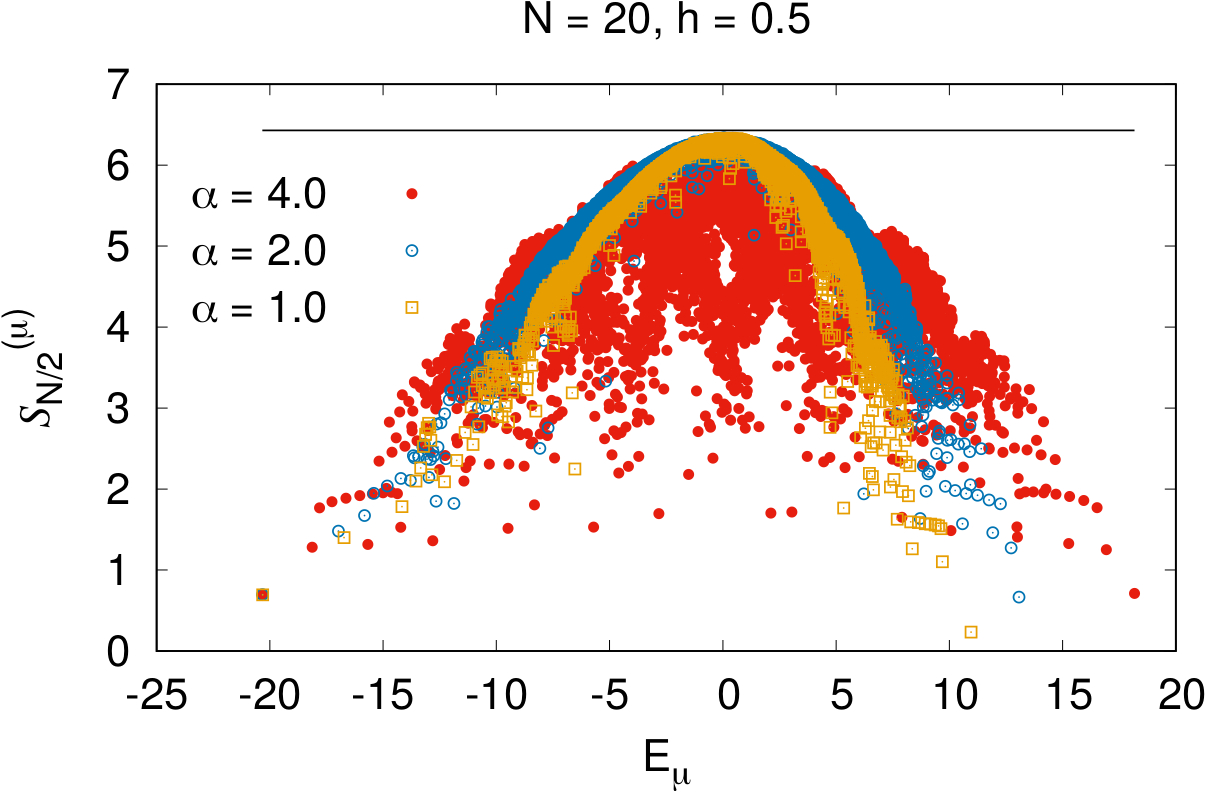}\put(90,50){(d)}\end{overpic}\\
   \end{tabular}
  \end{center}
 \caption{Half-system entanglement entropy of the eigenstates, scatter plot of $S_{N/2}^\mu$ versus $E_\mu$ for different values of the parameters. We consider $N=20$ and $h=0.1$ in panels (a),~(b) and $h=0.5$ in panels (c),~(d). The horizontal lines correspond to the Page value at $N=20$, the value corresponding to a fully random state~\cite{note}.}
    \label{correl2:fig}
\end{figure*}

Increasing $\alpha$ the multiplet structure disappears, first at higher, then at lower energy densities, as one can see in Fig.~\ref{correl2:fig}~(a) and~(c) already for $\alpha=0.5$ and $\alpha=0.75$. The scatter plot for $\alpha=2$ and $h=0.1$ [Fig.~\ref{correl2:fig}~(b)] is remarkable. Here the scatter plot {looks fuzzy and loses the smoothness typical of ETH}. For this value of $h$, $\alpha=2$ corresponds to a minimum in the level spacing ratio (see Fig.~\ref{correl1:fig}~(a) ). 

 Let us move to quantify the smoothness of the entanglement-entropy curves.
{Considering $S_{N/2}^{(\mu)}$, we wish to characterize its eigenstate to eigenstate fluctuations. In ETH these fluctuations should be smaller compared to other contexts, because $S_{N/2}^{(\mu)}$ should resemble the microcanonical curve, smooth in $E_\mu$. In order to quantify the fluctuations we consider 
\begin{align} \label{memma:eqn}
  \mathcal{M}&\equiv\frac{1}{\mathcal{N}_S-1}\sum_{\mu=1}^{\mathcal{N}_S-1}|S_{N/2}^{(\mu+1)}-S_{N/2}^{(\mu)}|\,,\nonumber\\
\end{align}
%
Here, $\ket{\varphi_\mu}$ and $\ket{\varphi_{\mu+1}}$ are ``nearby eigenstates''~\cite{Pal_PhysRevB10} with the $E_\mu$ and $E_{\mu+1}$ in \textit{increasing order}.  (unique for $\alpha>0$ and inside $\mathcal{H}_S$, where there are no degeneracies.) A quantity similar to $\mathcal{M}$ was introduced in~\cite{Pal_PhysRevB10} in the disordered Heisenberg chain taking instead of $S_{N/2}^{(\mu)}$ the local magnetizations. In case of {a system obeying} ETH, $\mathcal{M}$ is expected to {exhibit a rapid} decay upon increasing system size $N$.

We plot $\mathcal{M}$ versus $N$ in Fig.~\ref{seren:fig}. We compare with the case of the $\alpha\to\infty$ (nearest-neighbour) Ising model in transverse field in Fig.~\ref{seren:fig}~(b) and~(c). The nearest-neighbour model is integrable~\cite{Pfeuty}, and, consistently with that, the value $\mathcal{M}$ stays more or less constant with the size $N$. On the opposite, in the long-range model Eq.~\eqref{model:eqn}, $\mathcal{M}$ clearly decreases with $N$ for most of the considered values of $\alpha$. We emphasize that this occurs for the small values of $\alpha$, but we cannot tell if this is due to the entanglement-entropy curves getting smoother inside the multiplets or to the fact that the multiplets tend to merge with each other for increasing $N$. 

We see that there is a close correspondence between the decay of $\mathcal{M}$ with $N$ and the Wigner-Dyson value of the level spacing ratio (see Fig.~\ref{correl1:fig}). Indeed, the only conditions where we see something different from a decrease of $\mathcal{M}$ with $N$ in Fig.~\ref{seren:fig} correspond to values of $\alpha$ where the average level spacing ratio has not yet attained the Wigner-Dyson value. This is true for $\alpha=8$ [Fig.~\ref{seren:fig}~(b),~(c)] and, as we have argued in Sec.~\ref{spectrum:sec}, this is most probably a finite-size effect. This is also true for $h=0.1$ and $\alpha=2,\,2.25$ [Fig.~\ref{seren:fig}.~(b)]. The effect is very strong for $\alpha=2$, again suggesting a connection with the integrability of other $\alpha=2$ long-range spin chain models. 

\begin{figure}[h!]
  \begin{center}
   \begin{tabular}{c}
     \begin{overpic}[width=80mm]{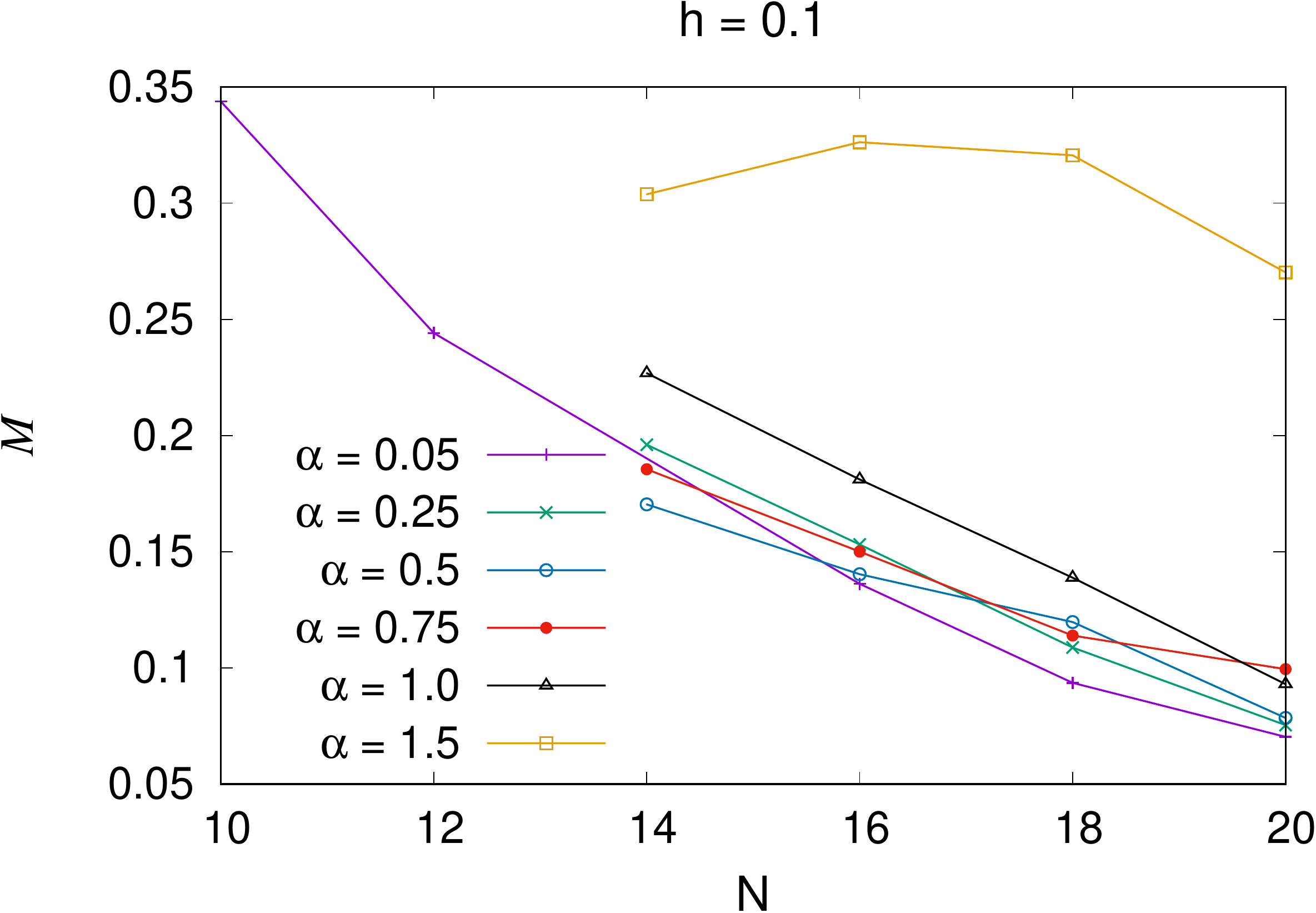}\put(20,43){(a)}\end{overpic}\\
     \\
     \begin{overpic}[width=80mm]{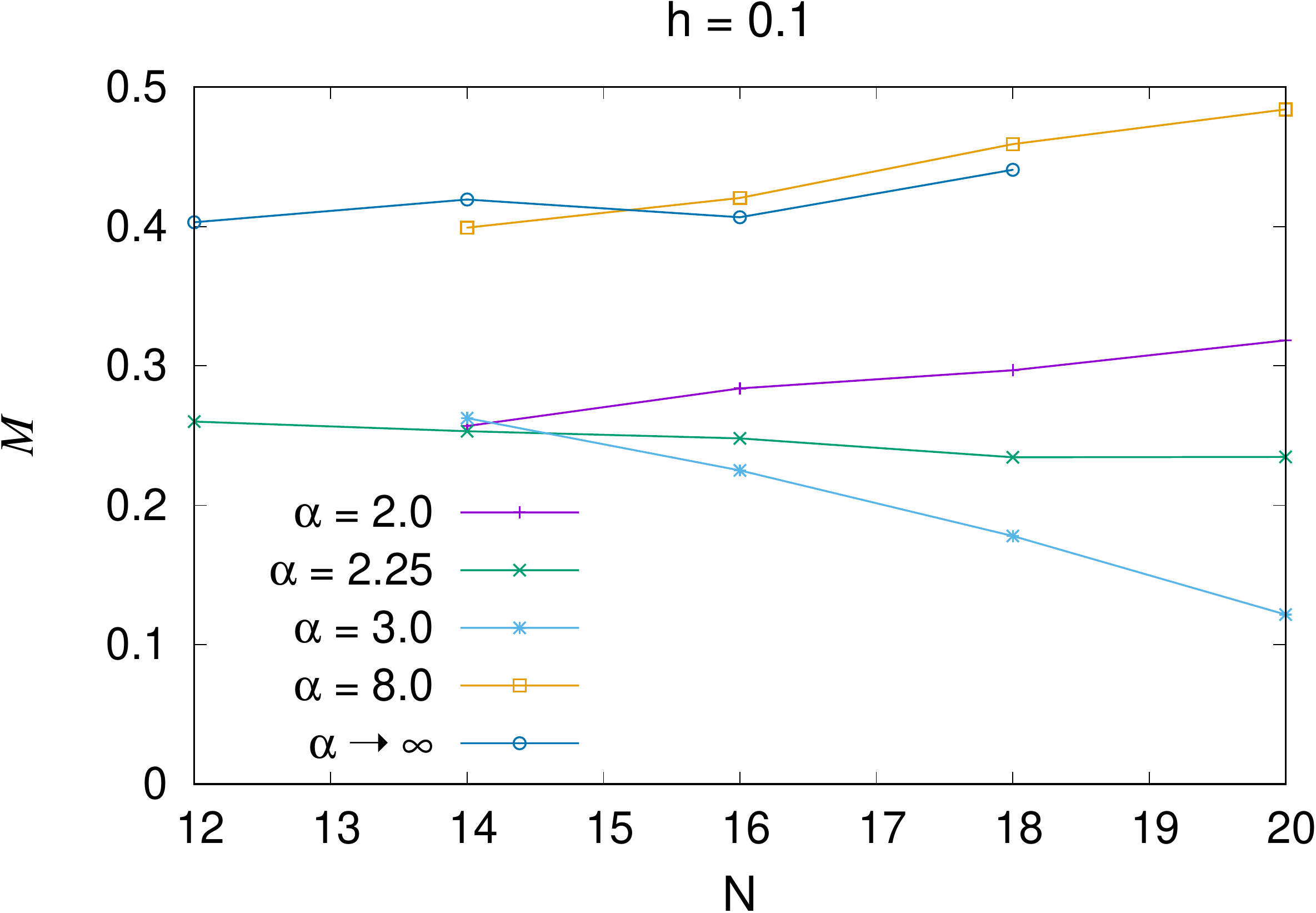}\put(18,43){(b)}\end{overpic}\\
     \\
     \begin{overpic}[width=80mm]{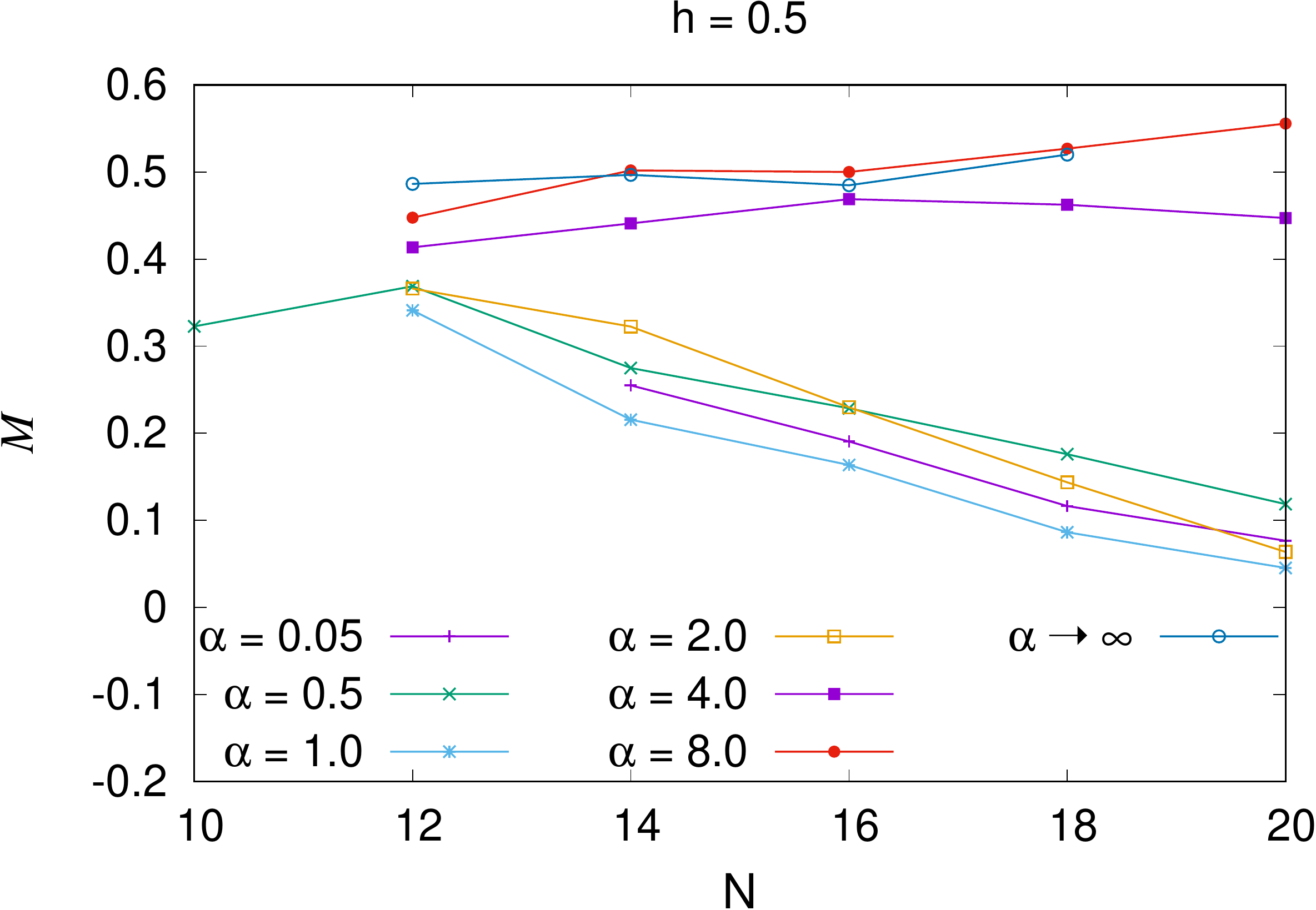}\put(20,50){(c)}\end{overpic}\\
   \end{tabular}
  \end{center}
 \caption{Plot of $\mathcal{M}$ versus $N$ for different values of $\alpha$ and $h$ and, for comparison, the Ising integrable $\alpha\to\infty$ nearest-neighbour model.}
    \label{seren:fig}
\end{figure}

{Another quantitative analysis relevant for the study of ETH is the comparison with the Page value. ETH eigenstates with the largest entanglement are expected to approach the so-called Page value~\cite{Page_PRL} upon increasing the system size $N$ (the Page value corresponds to the entanglement entropy of a fully-random state~\cite{note}).
We want to quantitatively probe this fact and consider the following two quantities introduced in~\cite{russomanno2020nonergodic}.}
The first one is defined as
\begin{equation} \label{lambda:eqn}
  \Lambda_S(N) = \frac{1}{\mathcal{N}_S}\sum_\mu\log\left(|S^{\rm (Page)}_{N/2}-S_{N/2}^{(\mu)}|\right)\,.
\end{equation}
The rationale is that the logarithm overweights the smallest values of the argument and the high-entropy states -- corresponding to the smallest values of the difference in the argument -- give the strongest contribution to the average. If the highest-entropy states tend to the Page value, $\Lambda_S(N)$ takes more and more negative values.

In order to define the second quantity, we need to first define the integer number $1\leq \mu^*\leq\dim\mathcal{H}_S$ as the value of $\mu$ such that the quantity $|S^{\rm (Page)}_{N/2}-S_{N/2}^{({\mu^*})}|$ is minimum over $\mu$. Restricting the average of the entanglement entropy to states around the energy $E_{\mu^*}$, we focus on the highest entropy states, the ones nearest to the Page value. More formally, if we term the width of the energy spectrum as $\Delta E(N)=\max_\mu(E_{\mu})-\min_\mu (E_{\mu})$, we restrict the sum to the states with eigenenergy $E_\mu\in[E_{\mu^*}-\frac{f}{2}\Delta E(N),E_{\mu^*}+\frac{f}{2}\Delta E(N)]$ (call their number $\mathcal{N}_f$). In this way we can define
\begin{equation} \label{S:eqn}
  \mean{S_{N/2}}_f=\frac{1}{\mathcal{N}_f}\sum_{\mu\,{\rm s.t.}\,E_\mu\in[E_{\mu^*}-\frac{f}{2}\Delta E(N),E_{\mu^*}+\frac{f}{2}\Delta E(N)]}S_{N/2}^{(\mu)}\,.
\end{equation}
We choose $f=0.2$, so that the sum is restricted around the state with entropy nearest to the Page value, that's to say to the infinite-temperature value. If $\Lambda_S(N)$ and $(S^{\rm (Page)}_{N/2}-\mean{S_{N/2}}_f)/N$ get smaller, the system becomes more ETH.

\begin{figure*}
  \begin{center}
   \begin{tabular}{cc}
     \hspace{-1cm}\begin{overpic}[width=80mm]{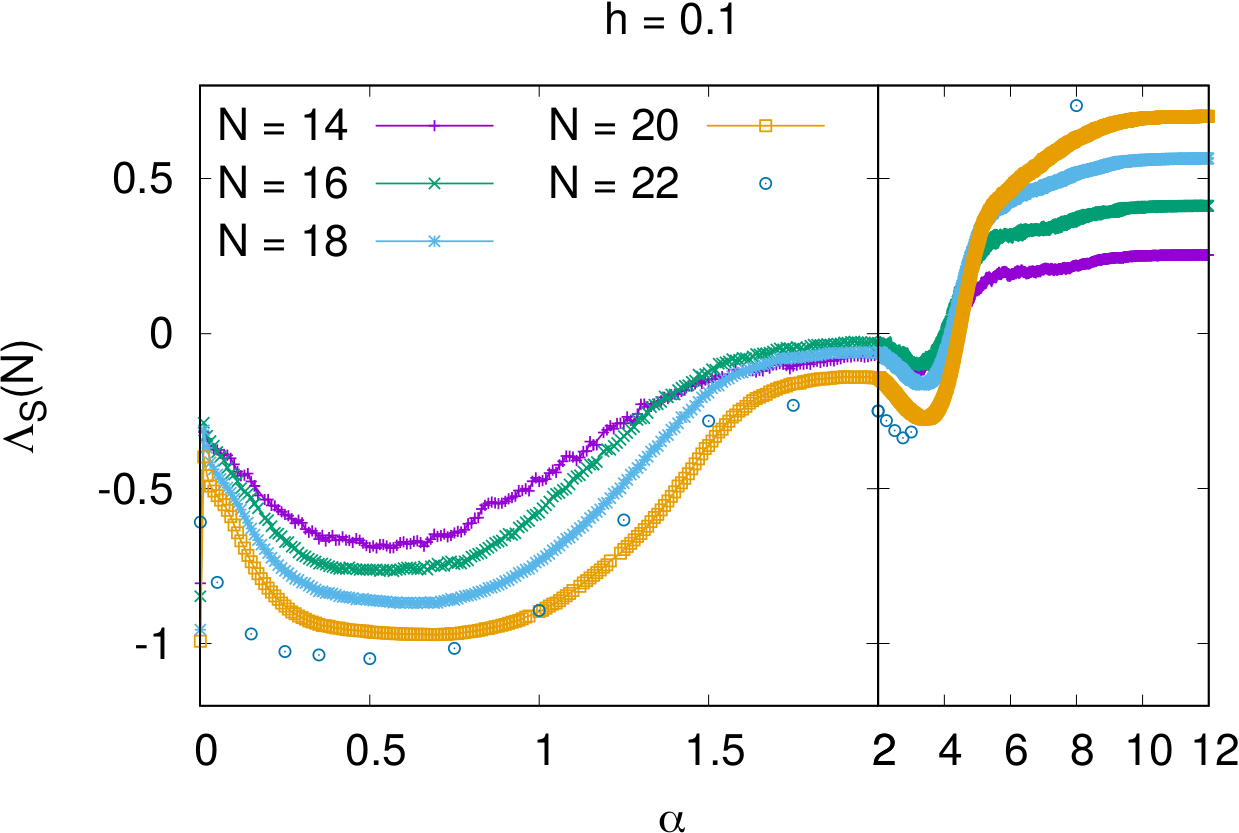}\put(87,15){(a)}\end{overpic} &
     \begin{overpic}[width=80mm]{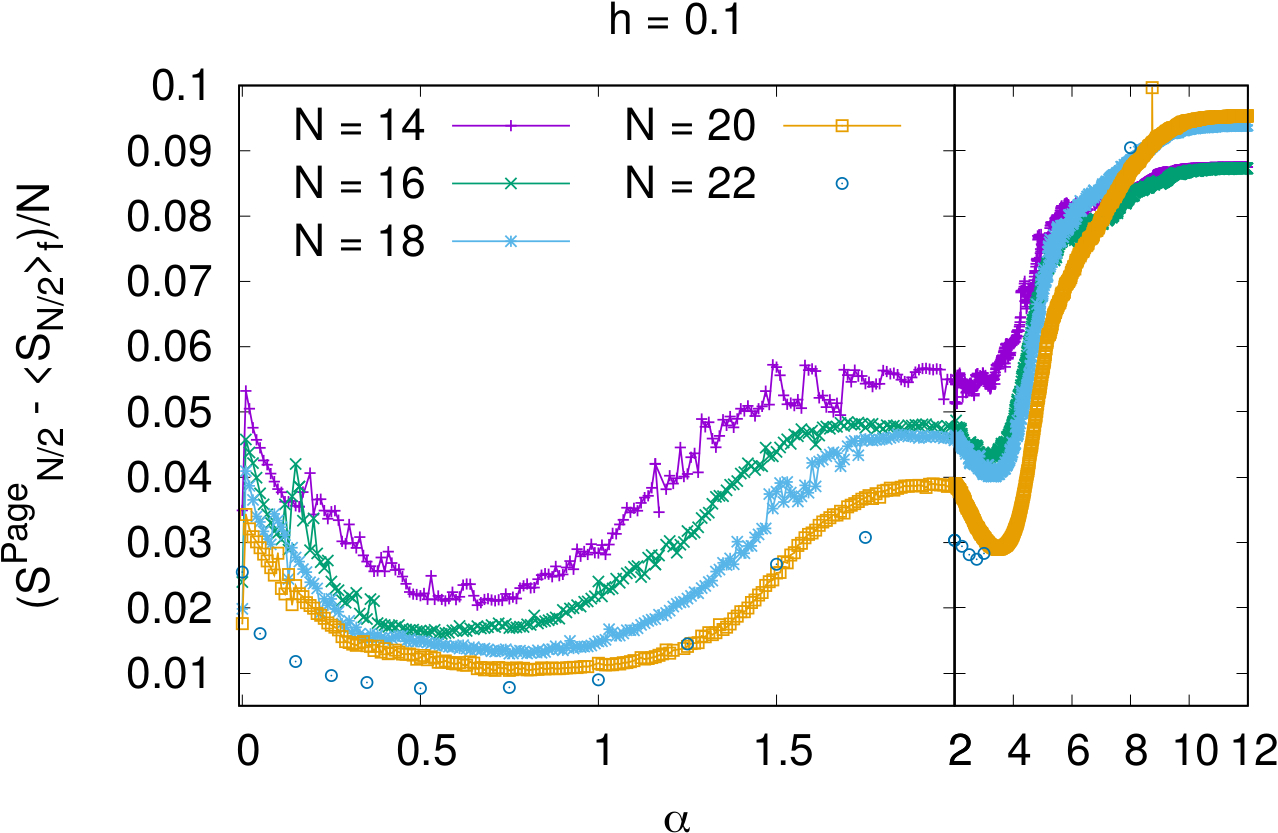}\put(87,15){(b)}\end{overpic} \\
     \\
     \hspace{-1cm}\begin{overpic}[width=80mm]{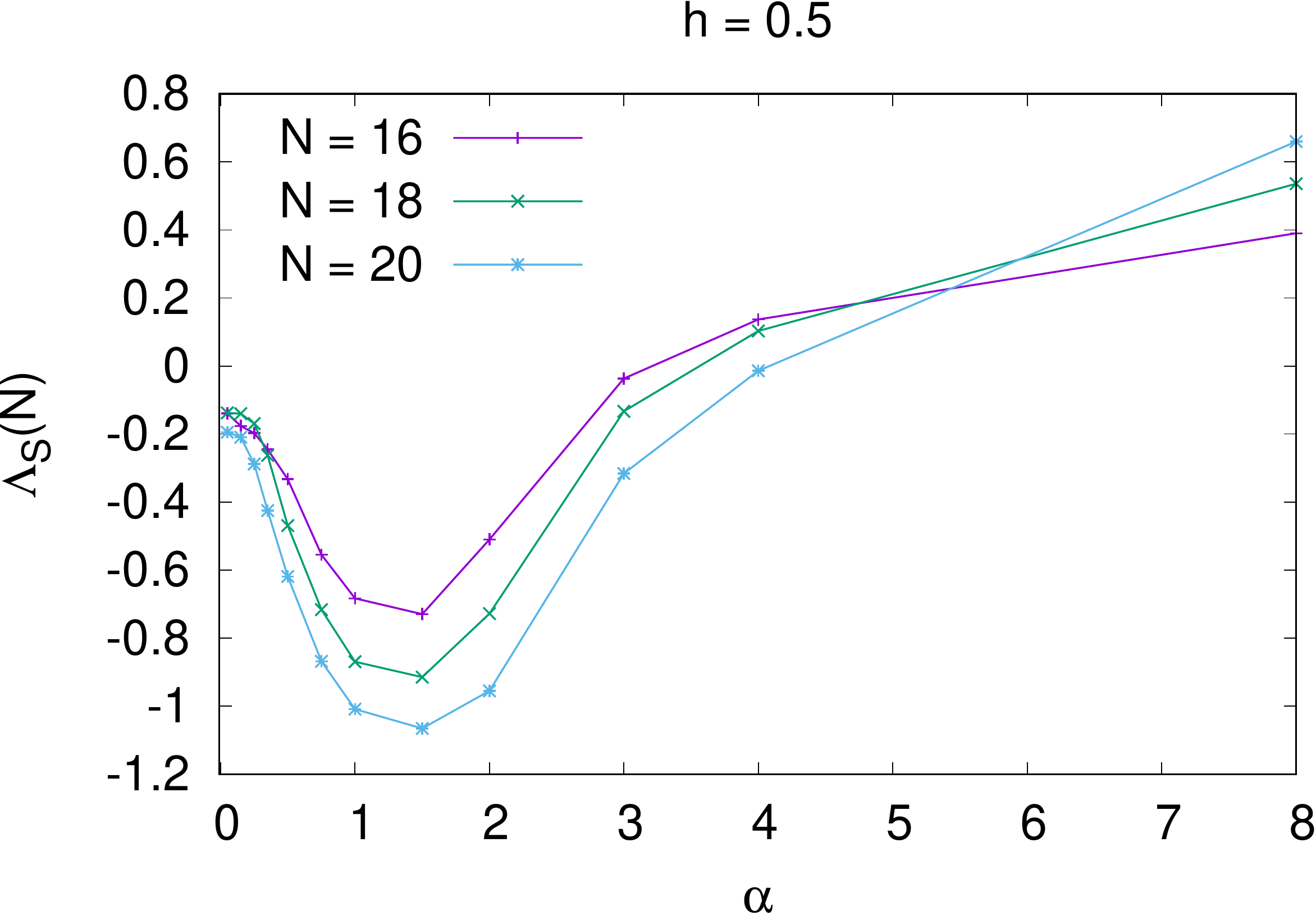}\put(87,15){(c)}\end{overpic} &
     \begin{overpic}[width=80mm]{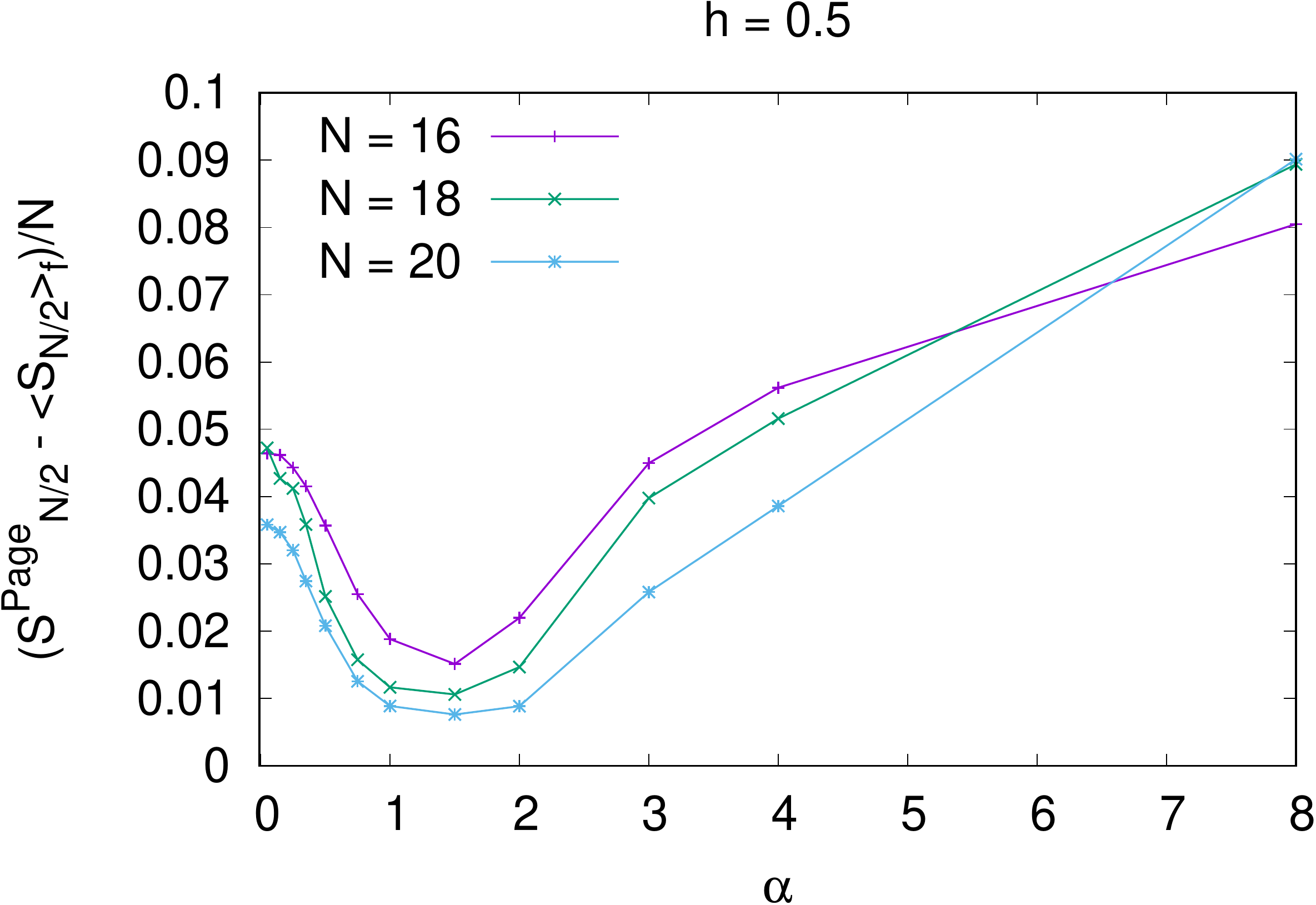}\put(87,15){(d)}\end{overpic}
   \end{tabular}
  \end{center}
 \caption{Plot of the quantities $\Lambda_S(N)$ [Eq.~\eqref{lambda:eqn} -- panels~(a),~(c)] and $(S^{\rm (Page)}_{N/2}-\mean{S_{N/2}}_f)/N$ [Eq.~\eqref{lambda:eqn} -- panels~(b),~(d)  -- $f=0.1$] versus $J$ for different values of $N$. $h=0.1$ in panels~(a),~(b) and $h=0.5$ in panels~(c),~(d). {For $N=22$ we consider 14000 randomly chosen eigenstates, in the other cases all the spectrum.}}
    \label{plots_entro:fig}
\end{figure*}
We report the results for $\Lambda_S(N)$ versus $\alpha$ for different values of $N$ in Fig.~\ref{plots_entro:fig}~(a),~(c), and those for $(S^{\rm (Page)}_{N/2}-\mean{S_{N/2}}_f)/N$ in Fig.~\ref{plots_entro:fig}~(b),~(d). The steady decrease with $N$ for $h=0.5$ suggest a tendency to ETH for increasing system size. The largest-$\alpha$ crossing point between curves with nearby values of $N$ tends to shift right for increasing $N$. The increase in $N$ for large $\alpha$ is therefore a finite-size effect. Results for $h=0.1$, on the opposite, are not that conclusive. Although the behavior at small and large $\alpha$ is similar to the $h=0.5$ case, we find an interval of $\alpha$ ($\alpha\in[1,1.5]$) where both the considered quantities seem to saturate with $N$. Quite remarkably, in this interval of $\alpha$ the average level spacing ratio is significantly different from the Wigner-Dyson value [see Fig.~\ref{correl1:fig}~(a)] and probably finite-size effects are too strong.
\section{Hilbert-Schmidt distance from the infinite-range model} \label{hilbers:sec}

The Hilbert-Schmidt  distance is an operator distance used in quantum information~\cite{PhysRevA.100.022103,PhysRevA.102.012409} and is defined by the norm
	$\left\lVert \hat{O} \right\rVert_{HS} = \sqrt{ \Tr\left( \hat{O}^\dag \hat{O} \right)}$.
We are going to show that the Hilbert-Schmidt distance of the Hamiltonian at $\alpha>0$ from the infinite-range Hamiltonian at $\alpha=0$  increases linearly with $\alpha$ when $\alpha$ is small. 

{We} consider the Hamiltonian Eq.~\eqref{model:eqn},  and we want to quantify the Hilbert-Schmidt distance of $\hat{H}^{(\alpha)}$ from its infinite-range $\alpha=0$ counterpart $\hat{H}^{(0)}$. We define the distance as
\begin{equation}\label{distH:eqn}
	d(\alpha, N) = \left\lVert\Delta H(\alpha, N)\right\rVert_{HS} = \sqrt{\text{Tr}\left[ \left(\Delta \hat{H}(\alpha, N)\right)^2 \right]},
\end{equation}
with $\Delta \hat{H}(\alpha,N) \equiv \hat{H}^{(\alpha)} - \hat{H}^{(0)}$ independent of $h$. Note that for an Hermitian operator $\hat{O}$ with eigenvalues $\lambda_j$, $\lVert O \rVert_{HS} = \sqrt{\sum_j \lambda_j^2}$.
To compute $d(\alpha,N)$, we write
\begin{equation}
	\Delta \hat{H}(\alpha, N) = \sum_{i,j,\,i\neq j}^N J_{i,j}'(\alpha) \hat{\sigma}_i^z \hat{\sigma}_j^z,
\end{equation}
where 	$J_{i,j}' = \frac{1}{N(\alpha) D_{i,\,j}^\alpha} - \frac{1}{N(0)}$. Then
\begin{align}\label{dH2:eqn}
	\left[\Delta \hat{H}(\alpha, N)\right]^2 &= \sum_{i,j,\,i\neq j}^N \left[J_{i,j}'(\alpha)\right]^2 + \sum_{\text{distinct } i, j, k}^N (\cdots) \hat{\sigma}_i^z \hat{\sigma}_j^z&\nonumber\\
  &+ \sum_{\text{distinct } i, j, k, l}^N (\cdots) \hat{\sigma}_i^z \hat{\sigma}_j^z \hat{\sigma}_\mu^z \hat{\sigma}_l^z&\,.
\end{align}
Taking the trace, all term but the first one vanish, so that
\begin{equation}\label{hilbdist:eqn}
	d(\alpha, N) = 2^{N/2}\sqrt{\sum_{i,j,\,i\neq j}^N \left[J_{i,j}'(\alpha)\right]^2}\,.
\end{equation}
We numerically compute this quantity for various values of $N$ and report it versus $\alpha$ in Fig.~\ref{fig}. We  clearly see that it increases linearly in $\alpha$ for small $\alpha$. 

\begin{figure}
  \begin{center}
    \begin{tabular}{c}
        \\
      \begin{overpic}[width=80mm]{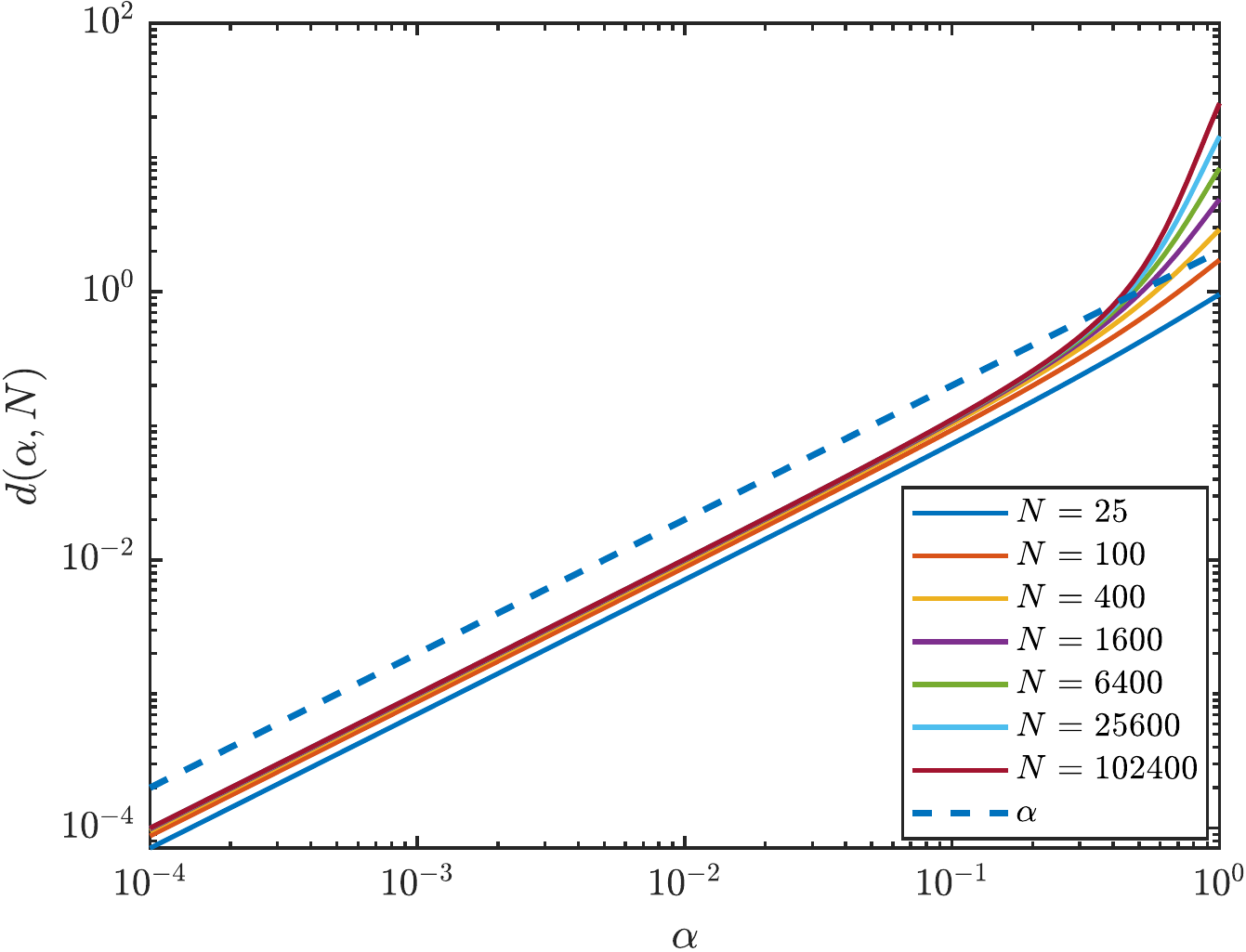}\put(3,51){\pgfbox[left,top]{\begin{pgfrotateby}{\pgfdegree{90}}$/2^{N/2}$\end{pgfrotateby}}}\end{overpic}
    \end{tabular}
	\caption{$d(\alpha, N)/2^{N/2}$ versus $\alpha$ for different values of $N$. Notice the linear increase with $\alpha$. \label{fig}}
  \end{center}
\end{figure} 
We strongly remark that, for $\alpha<1$, $d(\alpha, N)/2^{N/2}$ fast saturates to a constant when $N$ is increased. This point is crucial: The fact that $d(\alpha, N)/2^{N/2}$ is asymptotically constant with $N$ is at the root of our argument in Sec.~\ref{randma:sec}. This result can be seen in Fig.~\ref{fig} and can also be analytically checked in the large-$N$ limit, by using translational invariance and writing approximately
%
  $d(\alpha, N) \simeq 2^{N/2}\sqrt{2N\sum_{l>1}^{N/2} \left[\frac{1}{N(\alpha) }\frac{1}{l^\alpha} - \frac{1}{N(0)}\right]^2}$,
%
and then using the asymptotic behaviours $N(0)=N$, $N(\alpha)\sim N^{1-\alpha}$, $\sum_{l>1}^{N/2} \frac{1}{l^\alpha}\sim N^{1-\alpha}$.
%

\begin{thebibliography}{98}
\bibitem{nothak} The operator behaving as a random matrix when restricted to a multiplet, and the one providing the multiplet spectral width, can be indifferently $\Delta \hat{H}(\alpha,N)$ or $\hat{H}^{(\alpha)}$ because $\alpha=0$-degeneracy {implies that $\hat{H}^{(0)}$ restricted to the multiplet is proportional to the identity.}
\bibitem{notS} Remember that for each value of $S$ there are $2S+1$ distinct multiplets, as many as the eigenvalues of the total spin along $z$.
\bibitem{note_w} We use the random-matrix estimate for $w(N,S)$ Eq.~\eqref{w:eqn} for all the multiplets. This is an approximation
because the random-matrix assumption might not hold for multiplets near the edges of the spectrum. Nevertheless this is a good
approximation, because the role of states at the edges of the spectrum becomes negligible in the large-$N$ limit.
\end{thebibliography}
%
%
\end{document}